\documentclass[fleqn,usenatbib,useAMS]{mnras}

\usepackage{graphicx}	
\usepackage{amsmath}	
\usepackage{amssymb}	
\usepackage{multicol}        
\usepackage{bm}		
\usepackage{pdflscape}	
\usepackage{algorithm}
\usepackage{algpseudocode}

\usepackage{subfigure}
\usepackage{multirow}

\usepackage[T1]{fontenc}
\usepackage{ae,aecompl}

\usepackage{newtxtext,newtxmath}

\title[]{An investigation on the factors affecting machine learning classifications in $\gamma$-ray astronomy}

\author[S. Luo et al.]{
Shengda Luo,$^{1}$
Alex P. Leung, $^{1}$
C. Y. Hui,$^{2}$\thanks{E-mail: huichungyue@gmail.com, cyhui@cnu.ac.kr}
K.L. Li,$^{2,3,4}$
\\
$^{1}$Faculty of Information Technology, Macau University of Science and Technology, Avenida Wai Long, Taipa, Macau\\
$^{2}$Department of Astronomy and Space Science, Chungnam National University, Daejeon 34134, Korea\\
$^{3}$Department of Physics, UNIST, Ulsan 44919, Korea\\
$^{4}$Institute of Astronomy, National Tsing Hua University, Hsinchu, 30013, Taiwan\\
}

\date{Last updated 2015 May 22; in original form 2013 September 5}

\pubyear{2015}

\begin{document}
\label{firstpage}
\pagerange{\pageref{firstpage}--\pageref{lastpage}}
\maketitle

\begin{abstract}
We have investigated a number of factors that can have significant impacts on the classification performance of $\gamma$-ray sources detected by {\it Fermi} Large Area Telescope (LAT) with machine learning techniques. We show that a framework of automatic feature selection can construct a simple model with a small set of features which yields better performance over previous results. Secondly, because of the small sample size of the training/test sets of certain classes in $\gamma$-ray, nested re-sampling and cross-validations are suggested for quantifying the statistical fluctuations of the quoted accuracy. We have also constructed a test set by cross-matching the identified active galactic nuclei (AGNs) and the pulsars (PSRs) in the {\it Fermi} LAT eight-year point source catalog (4FGL) with those unidentified sources in the previous 3$^{\rm rd}$  {\it Fermi} LAT Source Catalog (3FGL). Using this cross-matched set, we show that some features used for building classification model with the identified source can suffer from the problem of covariate shift, which can be a result of various observational effects. This can possibly hamper the actual performance when one applies such model in classifying unidentified sources. Using our framework, both AGN/PSR and young pulsar (YNG)/millisecond pulsar (MSP) classifiers are automatically updated with the new features and the enlarged training samples in 4FGL catalog incorporated. Using a two-layer model with these updated classifiers, we have selected 20 promising MSP candidates with confidence scores $>98\%$ from the unidentified sources in 4FGL catalog which can provide inputs for a multiwavelength identification campaign.
\end{abstract}





\section{Introduction}
The advancements of astronomical instrumentation, large-scales surveys and the policy of open data access have led 
us into the midst of a revolution of data science. This makes the {\it knowledge discovery in databases} (KDD) become 
feasible \citep{ball2010}. However, in order to fully harness the power of the deluge of data, one has to employ
the techniques of machine learning and data mining. This is envisioned as {\it the fourth paradigm} in astronomy 
\citep{bell2009}. \footnote{The other three traditional paradigms are: theory, observation and computer simulation}. 

Machine learning has been applied in classifying objects in different wavelengths 
\citep[e.g.][]{farrell2015,buisson2015,miller2015,mirabal2012,parkinson2016classification} 
In this work, we will focus on the $\gamma-$ray regime. 

The successful launch of \emph{Fermi} Gamma-ray Space Telescope in 2008 has brought us into a golden era of $\gamma-$ray 
astronomy \citep[see][for a recent review]{hui2018}. 
Its large field-of-view enables it to continuously survey the whole sky every $\sim3$~hrs. Together with the
high sensitivity of the Large Area Telescope (LAT) onboard, both the volume and the quality of the data collected in the 
energy range from 100~MeV to 300~GeV are unprecedentedly high. With only three months of observations, LAT has already 
detected 205 sources with statistical significance $>10\sigma$ \citep{abdo2009}. For comparison, {\it Fermi}'s 
predecessor, the Energetic Gamma-Ray Experiment Telescope (EGRET) on the {\it Compton} Gamma-Ray Observatory have 
only detected 271 $\gamma-$ray sources in $\sim4.5$~years \citep{hartman1999}. 

{\it Fermi} has revolutionized $\gamma-$ray astronomy by significantly enlarging the
$\gamma-$ray source population, as well as 
discovering a number of new classes of $\gamma-$ray objects, 
such as globular clusters, millisecond pulsars, radio-quiet pulsars, classical novae.
In the third {\it Fermi} LAT source catalog 
\cite[3FGL,][]{acero2015}, over 3000 sources belong to $\sim20$ different $\gamma-$ray object classes 
have been detected based on the first four year data. About 
$\sim2/3$ of the 3FGL sources have been identified as pulsars or active galactic nuclei, which are the two 
largest broad classes. On the other hand, there are about $\sim1000$ sources in 3FGL have not yet been identified 
unambiguously. This provides us with a considerable data space to explore.  

Various identification campaigns have been launched so as to unveil the nature of these unidentified $\gamma-$ray 
sources. A number of investigations have utilized the conventional analysis to classify these sources
\citep[e.g.][]{hui2015}, which requires prior knowledge of $\gamma-$ray source properties in different classes. 
On the other hand, automatic classification techniques have also been experimented 
\citep{ackermann2012,mirabal2012,parkinson2016classification}. 
These techniques have a number of advantages over the conventional ones. 

For building classification model, the conventional methods require our current understanding of the emission properties of different $\gamma-$ray object classes. However,
owing to the relatively short history of $\gamma-$ray astronomy, our current understanding of different classes might be 
far from being complete. With new data continuously pouring in and dedicated investigations, the existing models are subjected 
to modifications and new characteristics of various classes can be established. This implies that both efficiency and 
accuracy of the conventional method in identifying the unclassified $\gamma-$ray sources are unlikely to be optimal. 

By employing machine learning techniques in automatic classification, instead of relying on prior knowledge, one let the data ``speak for themselves".
In this way, classification models are generated based on the respective inductive bias of machine learning methods and the current data without prior knowledge.
With an appropriate algorithm, attributes and patterns of the data that might be overlooked by human investigators can be possibly highlighted by machine. Furthermore, as the data volume increases 
monotonically, automatic algorithms definitely have advantages over the traditional approaches.   
Also, once the new data become available, they can be trivially incorporated in most algorithms and the model will be 
automatically updated. Therefore, in terms of the efficiency, accuracy and cost-effectiveness, the automatic classification 
schemes supersede the traditional analysis. 

A number of previous investigations have explored the machine learning techniques in classifying {\it Fermi} $\gamma-$ray sources. 
By applying classification tree and logistic regression to the first {\it Fermi} LAT catalog 
\citep[1FGL][]{abdo2010} that based on the first 11 months data, 
\citet{ackermann2012} have attained an average accuracy of $\sim80\%$ in their automatic scheme. \citet{mirabal2012} 
have trained a random forest classifier in predicting the class memberships for the unassociated object in the second {\it Fermi} 
LAT catalog \citep[2FGL][]{nolan2012} which is based on two years data. An average accuracy $\sim97\%$ is achieved in their 
work. \citet{parkinson2016classification}  have experimented various algorithms in classifying the sources in 3FGL 
catalog. Using the random forest technique, the authors obtained the best overall accuracy of $\sim96\%$ for pulsars (PSRs) and active Galactic nuclei (AGNs) classifications. On the other hand, an accuracy of $\sim90\%$ is attained for young pulsars (YNGs) and millisecond pulsars (MSPs) classification.

While there are a number of studies applying machine learning techniques in $\gamma$-ray astronomy, none of them have investigated the factors that can improve/hamper the performance in details. First, we aware that the power of automatic feature selection algorithms \citep[e.g.][]{guyon2003} has not been exploited. The feature selections in these previous works were somewhat relied on the knowledge of the human investigators. 
For example, the absolute Galactic latitude $|b|$ was not adopted as a feature in classifying 
the $\gamma-$ray objects into PSRs and AGNs as
this might introduce a bias against AGNs close to the Galactic plane and the pulsars away from it \citep{ackermann2012,mirabal2012,parkinson2016classification}. 
\citet{mirabal2012} have explicitly demonstrated that the classification 
accuracy is slightly lower when this parameter is included. However, for the classification between YNGs and MSPs, 
\citet{parkinson2016classification} have added $|b|$ as a predictor parameter. This ad hoc decision is simply based on our existing knowledge of the 
spatial distributions of MSPs and YNGs. 

Involving human decision in the classification in the previous works suggests that the power of the automatic approaches might not be fully harnessed.  

Another issue for $\gamma$-ray astronomy is the sample size of the training/test set is relatively small in comparison with other wavelength. For example, the YNG/MSP classification performed by \citet{parkinson2016classification} depending on a sample of only $<100$ labeled objects in each class. It is possible that fluctuations of any  performance metrics can be resulted from such small sample. However, such issue has not yet been properly addressed.

One basic assumption for any supervised learning is that the training and test data are drawn from the same distribution in the feature space. This assumption has not been examined in all previous studies, because the test sets and the training sets were constructed from the identified sources. However, in view of the selection effects (e.g. analysis of bright sources in details can be easier in pinpointing their natures), it is uncertain whether the model developed through the training with identified sources can legitimately be applied to classify unidentified sources. 

The objective of this paper is to investigate the impacts of these factors in $\gamma$-ray source classification, which is organized as follows: We gives an overview on the algorithm of automatic feature selection in Section ~\ref{sec_feature_selection} and describe how we incorporate this technique in our framework in Section ~\ref{sec_OurMethod}. In Section ~\ref{sec_metrics}, we describe the metrics that we have adopted in quantifying the model performance. In Section ~\ref{sec_3fgl_experiments}, we present the results of the classification of 3FGL sources by using our framework and compare with those reported by \citet{parkinson2016classification}. In Sections ~\ref{sec_covariate_shift} and ~\ref{sec_cross_matched}, we introduce the concept of covariate shift and examine its possible impacts on 3FGL classification. In Section ~\ref{sec_4fgl_msp}, with these factors taken into account, we have selected 20 promising MSP candidates from the unidentified sources in 8-year {\it Fermi} LAT point source \citep[4FGL][]{fermi2019} catalog for follow-up investigations in the future. And lastly, we summarize our work in Section ~\ref{sec_summar}.

\section{Feature Selection}
\label{sec_feature_selection}
\subsection{An Overview for Feature Selection}
In the context of machine learning, the term feature selection refers to the process that extracts an effective subset of features from the feature space for better classification performance and/or lower computational complexity. 
The goal of feature selection is to improve the performance of models built by machine learning techniques. 
Especially, it is important for improving model performance with a feature space in high dimensionality.

Among many existing methods for feature selection, we select Recursive Feature Elimination (RFE) in our work because of its good performance as well as its simplicity. 
However, other feature selection methods can also be very effective 
methods \citep{Sanchez2007,Nguyen2014,vergara2014,bennasar2015}.

\subsection{An Overview for Recursive Feature Elimination}
RFE is one of the best performing methods for feature selection \citep{Richert2013}. 
RFE is a backward selection method and, therefore, unimportant features are sequentially eliminated during a recursive process.
To select features, RFE has to be combined with a classification method producing the important scores.
It is worth to notice that the classification method combined with RFE in the feature selection stage can be a different method from the eventual classifier in the classification stage (see Algorithm~\ref{sec_OurMethod}). 
For example, it is possible that a combination of Random Forest with RFE for feature selection and the Support Vector Machine (SVM) can lead to an optimal classification performance.

Unimportant features are iteratively eliminated in RFE.
In the $i$-th iteration, first, the classification method combined with RFE is fitted to the remaining $N-i$ features, and then, the performance of the fitted model $M_{i}$ is evaluated.

Finally, the important scores of $N-i$ features are generated using the fitted model $M_{i}$, and the feature with the lowest important score is eliminated.
The remaining $N-i-1$ features are used for the next iteration. 
The classification method combined with RFE is fitted to the various subsets of $N$ features, 
and totally $N$ fitted models are generated.
Features used to build the fitted model with the best classification performance is the output of RFE.


\subsection{Recursive Feature Elimination with Random Forest}
In this work, Random Forest (RF) is used as the classification method combined with RFE.
RF is successfully used to deal with a lot of classification problems with high-dimensional data \citep{genuer2017random,gomes2017adaptive,belgiu2016random,ristin2015incremental}.
RF is one of the most famous bagging methods, and it aggregates the predictions from a large number of bootstrapped trees for classification.
When an object is predicted by an RF classifier, it takes the results from each classification tree with its weight into account. 
With bagging, this algorithm can provide a relatively high stability and give a more reliable classification rule. 

In RFE with RF, RF is iteratively fitted to the various subsets of $N$ features, and totally $N$ RF classifiers are generated.
$N$ denotes the number of features of input data.
In each iteration, the performance of the RF classifier and feature important scores used to build the classifier are computed.
The performance of the $N$ classifiers are compared, and the set of features $F$ used to build the classifier with the best performance is selected to be the best set of features.

\section{A proposed simple Automatic Framework}
\label{sec_OurMethod}
In this work, a
simple
framework with automatic feature selection is proposed to classify $\gamma$-ray sources automatically. 
The proposed framework (see Algorithm~\ref{alg_OurMeth}) consists of four stages:
\begin{enumerate}
\item
Preprocessing data, 
\item
Automatic feature selection, 
\item
Building the prediction models, and
\item
Classification/Prediction.
\end{enumerate}
The stages of the proposed framework are introduced in detail in the following subsections.
The feature selection method is only applied on the training set. Features are selected based on the best performance on the training set only.
It means that Algorithm~\ref{alg_rfe_rf} is only applied on the training set.
In Algorithm~\ref{alg_OurMeth}, Stage 1 to 3 are performed on the training data, and only Stage 4 is performed on the test data.

\begin{algorithm}[H] 
\caption{The Proposed Framework}
\label{alg_OurMeth}
\renewcommand{\algorithmicrequire}{\bf Input:}
\renewcommand{\algorithmicensure}{\bf Output:} 
\begin{algorithmic}
\Require 
The training set and test set; a threshold: $E$
\Ensure 
The prediction results of test set.
\State   \texttt{Stage 1 to 3 are performed on the training data}
\State    \texttt{ Stage 1: Preprocessing Data} %
	\State 1: Count the number of empty entries per feature
	\State 2: Remove features with more than $E$ empty entries
	\State 3: The values of empty entries in the features with less than $E$ empty entries are filled with the mean of the features
	\State 4: (Optional) Clean features 
	\State 5: (Optional) Adding manual features using prior knowledge
\State  \texttt{ Stage 2: Feature Selection} 
	\State 6: Select sources' features by the variant of RFE with RF (see Algorithm ~\ref{alg_rfe_rf})
\State   \texttt{ Stage 3: Building the prediction models} 
	\State 7: Build seven prediction models based on various machine learning methods
	\State 8: Predict the training set using prediction models
	\State 9: Calculate ``Train ROC" curves using the predicting results of the training set
\State   \texttt{ Stage 4: Predicting} 
\State   \texttt{Only Stage 4 is performed on the test data} 
	\State 10: Predict the test set using the prediction models
	\State 11: Calculate ``Test ROC" curves using the predicting results of the test set
\end{algorithmic} 
\end{algorithm}

    \subsection{Data Preprocessing}
The purpose of the data preprocessing stage is to deal with missing data and to add more useful information with extra manual features to the $\gamma-$ray sources before subsequent stages. 
All data preprocessing steps in this stage are shown in Algorithm~\ref{alg_OurMeth} where $E$ is the threshold defined manually and $E$ is chosen to be 95\% in our experiments.

To handle missing data, any feature with more than $E$ empty entries are removed because there are too many sources with insufficient information for that particular parameter. 
For the features which have empty entries less than $E$, the empty entries are filled with the mean value of respective feature values.
For example, we assume that there is a four-dimensional feature shown as a vector $\left [ 1,2,miss,3 \right ]^{T}$.
Only one empty entry in this four-dimensional feature, and it with less than $E$ (here is 95\%) empty entries.
The value of the empty entry is filled with the mean of the four-dimensional feature, and the vector of feature is shown as $\left [ 1,2,2,3 \right ]^{T}$ after handling missing data (see the third step of Algorithm~\ref{alg_OurMeth}).

This is the general data preprocessing procedure adopted in this work with minimal modification of the data.
However, for comparing our results with those of \cite{parkinson2016classification} in classifying 3FGL sources, 
we add exactly the same extra manual parameters (e.g. hardness ratios) and apply the same cleaning methods used in \cite{parkinson2016classification}. 
Hence, the procedures of cleaning and adding manual parameters for the 3FGL catalog are exactly 
the same as the procedures used in \cite{parkinson2016classification}. 

    \subsection{Feature Selection}
    \label{subsec_fs}
In Algorithm~\ref{alg_OurMeth}, the feature selection method is only applied on the training set.
    
The output of any feature selection method is a set of $F$ features with low redundancy, which results in effective training for classifiers to achieve higher performance. 
A simplicity variant of the RFE with RF (see Algorithm~\ref{alg_rfe_rf}) is used for the feature selection stage in our framework (see Algorithm~\ref{alg_OurMeth}).
What motivates us to develop a simplicity variant of the RFE with RF is the question: why do we only trust the machine learning model and ignore why it made a correct prediction?
Motivated by Occam's Razor, one can pay for the simplicity with a small drop in predictive performance. 
A lot of feature selection techniques have been criticized for the lack of guarantees about their simplicity 
\citep{hastie03,johnstone09,loh12,lipton2016mythos,ribeiro2016model,guidotti2019survey}.
By trial and error, we accept a factor of 1.05 as the margin of error in the RMSE value to trade for a simpler model. 

As shown in Algorithm~\ref{alg_rfe_rf}, the simplicity variant is started with fitting RF to all features of input data, and then the less important features are removed one by one in a loop.
In the loop, RF is iteratively fitted to various subsets of $N$ features, and totally $N$ RF classifiers are generated.
In each iteration, the performance and important scores of an already fitted RF classifier are computed, and are recorded in a matrix $D$.
In Algorithm~\ref{alg_rfe_rf}, the root-mean-square error (RMSE) is used to evaluate the performance of the RF classifiers with a 10-fold cross-validation
on the training set.

Finally, after the loop, the output of Algorithm~\ref{alg_rfe_rf} is the set of features $F$ used to build the classifier with best trade-off between model prediction accuracy and simplicity.
Here, we consider that the best trade-off is to accept a factor of 1.05 as the margin of error in the RMSE value to trade for simplicity.
For example, assume that $N$ is equal to five, and $C_{1}$, $C_{2}$, $C_{3}$, $C_{4}$, $C_{5}$ are the five RF classifiers built by using 1, 2, 3, 4, 5 features.
The RMSE error of the five classifiers is 1.06, 1.04, 1.03, 1, 1.07 in order, and $C_{2}$ is the classifier with the best trade-off.

\begin{algorithm}[] 
\caption{Recursive Feature Elimination with Random Forest in the Proposed Framework.}
\label{alg_rfe_rf} 
\renewcommand{\algorithmicrequire}{\textbf{Input:}}
\renewcommand{\algorithmicensure}{\textbf{Output:}} 
\begin{algorithmic}[1] 
\Require 
The training set with $N$ features 
\Ensure 
$F$; $P$.
\State Train the RF classifier of RFE using the training set with $N$ features
\State Calculate the performance of the classifier (RF)
\State Calculate the importance scores of $N$ features
\State Record the performance and the features with their importance scores in $D$.
\For{i = 1 ... $N-2$,$N-1$}
	\State Eliminate the feature with the lowest importance score
	\State Train the RF classifier of RFE using the training set with remaining $N-i$ features
	\State Calculate the performance of the RF classifier
	\State Calculate the importance scores of $N-i$ remaining features
	\State Record the performance of the RF classifier and the $N-i$ remaining features with their importance scores in $D$.	
\EndFor
\State Obtain the performance profile, $P$, from $D$
\State Obtain the set of features, $F$, which are used to build the classifier with the best trade-off (see the second paragraph of Subsection~\ref{subsec_fs})
\end{algorithmic} 
\end{algorithm}

\subsection{Building the Prediction Models}
\label{sec_predcitionModel}
After feature selection with Algorithm~\ref{alg_rfe_rf}, we compare the performance of seven prediction models on the training set.
These models are built using the following methods: 
Random Forest (RF), Generalized Additive Models (GAM),
Logistic Regression (LR), Boosted Logistic Regression (Boost LR), 
Support Vector Machines (SVM), 
Decision Trees (DT) and Logistic Trees (LT).

During the process of building the prediction models, some parameters of various classifiers are tuned for optimizing their performances. 
Such parameters are automatically optimized by using a 10-fold cross-validation.
We use the same method (and the same code) as in  \cite{parkinson2016classification} to find the best classifier thresholds using training ROC curves (see Section ~\ref{sec_metrics} for a detailed description). And hence, if there is any difference between our results and 
\cite{parkinson2016classification}, it should come from the incorporation of our feature selection scheme. 

    \subsection{Predictions} 
At this stage, the test set is evaluated with the features from the aforementioned seven prediction models.
Different machine learning algorithms have their own characteristics and are suitable for different tasks.
RF, a widely used ensemble learning method, consists of a certain number of decision trees used as based learners \citep{breiman2001random,painsky2016cross,zhu2015reinforcement}. In the RF method, the prediction result is determined by voting of the decision trees trained using various subsets of features. 
GAM, a statistical model, is based on the assumption that the input data can be modeled linearly \citep{hastie2017generalized}. In this method, smooth functions are learned for predictions. 
LR, also a statistical model, draws on concepts from regression analyses \citep{hosmer2013applied}. This model is usually applied to the two-class classification problem and the probabilities of belonging to each class are calculated based on a set of independent features.
Boost LR, an ensemble learning method, draws on concepts from adaptive boosting (AdaBoost) \citep{friedman2000additive}. Boost LR is similarly to AdaBoost \citep{freund1999short} except Boost LR uses a negative log-likelihood loss function, instead of the exponential loss function used in AdaBoost.
SVM, traditionally a statistical machine learning method, is one of the most popular methods for two-class classification \citep{burges1998tutorial,hsu2002comparison,chang2011libsvm}. In SVM, a hyperplane is constructed for classification before the data in the high-dimensional space is mapped to the low-dimensional space.
DT, an inductive learning method, is a basic method for classification. It consists of two steps: 1.) the construction of tree and 2.) pruning \citep{quinlan2014c4,steinberg2009cart}.
LT is a method based on both LR and DT \citep{landwehr2005logistic}. 
The LT is constructed by using standard decision trees with logistic regression functions on the leaves. This method uses the LogitBoost algorithm to create a logistic regression function on the nodes of the tree.

\section{Performance Metrics}
\label{sec_metrics}
In order to completely evaluate the performance of the $\gamma$-ray classification methods, there are four performance metrics used in our experiments: the receiver operating characteristic (ROC) curve, the area under the curve (AUC), the average accuracy (see Equation~\ref{equ_aver_Acc}), and the standard error (see Equation~\ref{equ_sd}).

The ROC curve and AUC are two of the most popular evaluation methods for a binary classifier.
The curve is a plot of sensitivity against specificity.
The sensitivity of a ROC curve is the probability that a true positive is indeed classified as such by the model, while the specificity gives the probability of false alarm. 
Moreover, the sensitivity is also referred to as the true positive rate (TPR), and the specificity equals one minus the false positive rate (FPR). 
The AUC value is equal to the probability that a randomly chosen positive example is ranked higher than a randomly chosen negative example.
An ROC curve is obtained by calculating TPR against FPR at various thresholds. 
For each threshold, TPR and FPR are computed using true negatives (TN), false negatives (FN), true positives (TP), false positives (FP):
\begin{equation}
TPR = TP / \left( TP+FN \right),
\label{equ_TPR}
\end{equation}
\begin{equation}
FPR = FP/ \left( FP+TN \right).
\label{equ_FPR}
\end{equation}

The classification accuracy computed using TN, FN, TP, and FP: 
\begin{equation}
\label{equ_Acc}
{\rm Accuracy} = \left (TP + TN\right ) /\left ( TP + TN + NP +NF \right )
\end{equation}
is another important metric for quantifying the performance of classification models. 
Because of the randomness of algorithms (e.g. different bootstrapped trees in RF), although the same group of $\gamma$-ray sources are classified by the same machine learning models, experiment results show a little different.
In order to clearly evaluate the performance of $\gamma$-ray classification methods, each method is executed ten times with resampling by bootstrap, 
and the average over these ten accuracies (see Equation~\ref{equ_Acc}) is taken as the accuracy of the method (see Equation~\ref{equ_aver_Acc}).
Hence, Equation~\ref{equ_Acc} are re-defined as:
\begin{equation}
{\rm Accuracy} = \frac{1}{10}\sum_{i=1}^{10} \frac{TP_{i} + TN_{i}}{TP_{i} + TN_{i} + NP_{i} +NF_{i} },
\label{equ_aver_Acc}
\end{equation}
where $TP_{i}$, $TN_{i}$, $NP_{i}$ and $NF_{i}$ are generated the $i$-th time the classification method is run. 
In addition, the average accuracy are reported with the standard error obtained as:
\begin{equation}
{\rm S. E.} =(\frac{1}{10-1}\sum_{i=1}^{10}\frac{TP_{i} + TN_{i}}{TP_{i} + TN_{i} + NP_{i} +NF_{i}} - {\rm Accuracy})^{\frac{1}{2}}
\label{equ_sd}
\end{equation}
where ${Accuracy}$ here is computed by Equation~\ref{equ_aver_Acc}.

\section{Comparing the performance in classifying 3FGL sources with Saz Parkinson et al. (2016)}
\label{sec_3fgl_experiments}
We have performed a series of experiments to demonstrate the impact of including an automatic feature selection algorithm in classifying $\gamma-$ray sources in a catalog with a high-dimensional feature space. 
We intend to examine the impact of incorporating the procedure of automatic RFE with RF for classifying the same set of data by comparing our results with those of \cite{parkinson2016classification}. 
The features are automatically selected by a machine learning technique in our method, instead of manually selected as in the case of \cite{parkinson2016classification} \citep[see][Table 1]{parkinson2016classification}.
For comparing our results with previous results using the 3FGL catalog, classification methods and the way of preprocessing the data are exactly the same as those in \cite{parkinson2016classification}. 
The major improvement here in our experiments hence comes from the enhanced performance by selecting
an optimal feature set automatically.

For the motivations mentioned above, we follow \cite{parkinson2016classification} to divide the 3FGL catalog into binary classes for PSR/AGN and YNG/MSP classifications.
Our method (i.e. Algorithm~\ref{alg_OurMeth}) and that proposed by \citet{parkinson2016classification} are both evaluated using 3FGL catalog with the most updated source labels.
Currently, there are 1904 $\gamma$-ray labeled sources (166 PSRs and 1738 AGNs) which can be used in the PSRs/AGNs classification. 
And the PSRs can be further divided for YNGs/MSPs classification.
For PSRs/AGNs classification, the sample size is the same as that used by \cite{parkinson2016classification},
On the other hand, we have 13 more newly identified MSPs available for YNGs/MSPs classification than in case of \cite{parkinson2016classification}.
Additionally, we follow exactly the same procedures in preprocessing the data. This includes: (1) adding the 
points in the spectral energy density distribution (SED) in five different bands (Band 1. 0.1-0.3~GeV; Band 2. 0.3-1~GeV; Band 3. 1-3~GeV; Band 4. 3-10~GeV; 
Band 5. 10-100~GeV) and 
the derived hardness ratios in the feature space; (2) eliminating sources with missing values for any 
predicting features as adopted by \cite{parkinson2016classification}; (3) for each tested classifier, the data is randomly divided into the training set (70\%) and the test set (30\%).

\begin{figure}
\centering
\includegraphics[width=3in,height=2.5in]{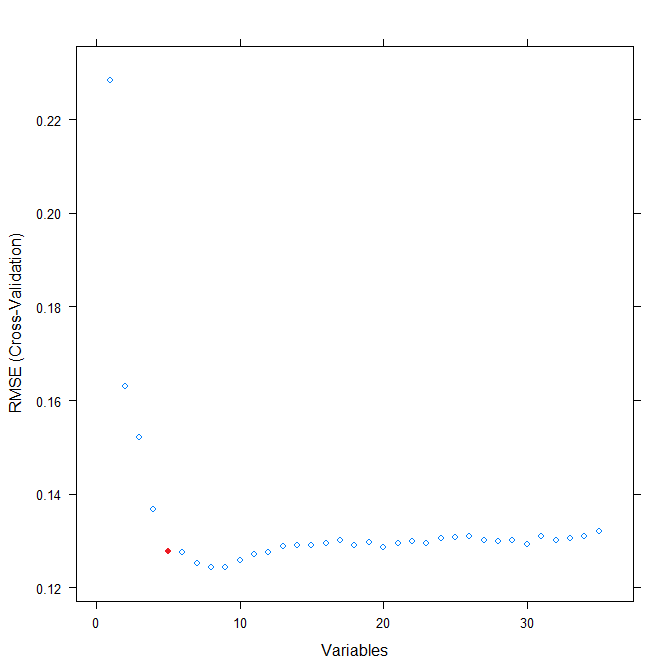}
\caption{The RMSE profile of AGN/PSR classification in the 3FGL catalog.
This profile is obtained with the training set. 
The classifier with the best trade-off between model prediction accuracy and model simplicity is trained with five features (the red point).}
\label{fig_profile_agn}
\end{figure}

\begin{figure}
\centering
\includegraphics[width=3in,height=2.5in]{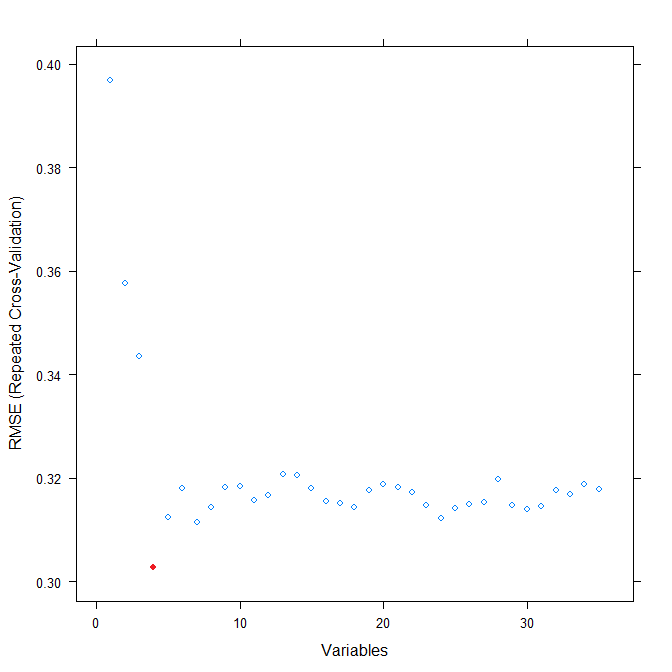}
\caption{The RMSE profile of MSP/YNG classification in the 3FGL catalog. 
This profile is obtained with the training set. 
The classifier with the best trade-off between model prediction accuracy and model simplicity is trained with four features (the red point).}
\label{fig_profile_yng}
\end{figure}

Starting with 35 features after data preprocessing, our feature selection method with the RFE technique (see Algorithm~\ref{alg_rfe_rf}) selects a subset of features for each task of classifying AGN/PSR and YNG/MSP.
For each task, we measure the feature selection performance with the RMSE against the number of selected features produced by the RFE with RF algorithm (see Algorithm~\ref{alg_rfe_rf}).
The feature selection performances for AGN/PSR and YNG/MSP classification tasks are shown in Figure~\ref{fig_profile_agn} and Figure~\ref{fig_profile_yng} respectively.
In our feature selection method, we accept a factor of 1.05 as the margin of error in the RMSE value to trade for simplicity.
In Figure~\ref{fig_profile_agn} and Figure~\ref{fig_profile_yng}, the red points denote the number of features used to build the classifiers with the best trade-off between model prediction accuracy and model interpretation.
Features selected by our method are ranked by the importance scores of features and summarized in Table~\ref{tab_3fgl_agnpsr_fea} and Table~\ref{tab_3fgl_mspyng_fea}.

In both classification tasks, it is shown that the prediction models in our framework are built with fewer features than that in \cite{parkinson2016classification} and hence our model is simpler.
Using Algorithm~\ref{alg_rfe_rf}, only five features (Table~\ref{tab_3fgl_agnpsr_fea}) are selected for the AGN/PSR classification task which is less than the set of nine features adopted by \cite{parkinson2016classification}. 
Also, using our method, a smaller set of features with four features (Table~\ref{tab_3fgl_mspyng_fea}) can lead to a better performance for YNG/MSP classification. Figure~\ref{fig_profile_yng} explicitly shows that this set of features gives rise to the global minimum in the RMSE profile.

For AGN/PSR classification, four selected features are related to the spectral shape in $\gamma-$ray. \texttt{Signif\_Curve} which depicts the curvature of the $\gamma-$ray spectrum receives the highest importance score. \texttt{Spectral\_Index} and the hardness ratio \texttt{hr45} are also determined by the $\gamma-$ray spectrum. For the uncertainty of integrated photon flux in $1-100$~GeV (\texttt{Unc\_Flux1000}), it can either reflect the quality of the spectral fit in this energy band 
(in which pulsars typically have exponential cut-off in their spectra) or as a proxy for the brightness in this band 
(see Section~\ref{sec_cross_matched} for further discussion). On the other hand, the other selected feature \texttt{Variability\_Index} enables one to differentiate PSR from AGN by their temporal behavior as the $\gamma-$ray emission from AGNs is typically highly variable..

In \cite{parkinson2016classification}, the Galactic latitude, \texttt{GLAT}, is manually removed in the AGN/PSR classification task and the same feature is manually chosen to be included in the YNG/MSP classification task based on the current knowledge of spatial distributions of YNG and MSP in the Milky Way. 

In our method, however, such feature is automatically included/excluded in classifying YNG/MSP and AGN/PSR. The selection is entirely based on the training data without any human intervention. 

On the other hand, the other three selected features for YNG/MSP classification are related to the uncertainty of flux at different bands. As aforementioned, these parameters reflect either the differences of the spectral steepness/curvature and/or the brightness between two classes. 

\begin{table}
\centering
\begin{tabular}{ c | c}
\hline
\hline
Features & Importance Scores \\
\hline
Signif\_Curve & 33.00 \\
\hline
Variability\_Index & 29.34 \\
\hline
Spectral\_Index & 24.97\\
\hline
Unc\_Flux1000 & 20.86 \\
\hline
hr45 & 19.47 \\
\hline
\end{tabular}
\caption{The rank of the features selected by Algorithm~\ref{alg_rfe_rf} for the AGN/PSR classification on the 3FGL training data. 
Please refer to Acero et al. (2015) for the physical meanings of these features.}
\label{tab_3fgl_agnpsr_fea}
\end{table}

\begin{table}
\centering
\begin{tabular}{ c | c}
\hline
\hline
Features & Importance Scores\\
\hline
Unc\_Flux1000 & 18.39 \\
\hline
Unc\_Energy\_Flux100 & 15.99 \\
\hline
Unc\_Flux\_Density &  9.94\\
\hline
GLAT &  8.87\\
\hline
\end{tabular}
\caption{The rank of the features selected by Algorithm~\ref{alg_rfe_rf} for the YNG/MSP classification on the 3FGL training data. 
Please refer to Acero et al. (2015) for the physical meanings of these features.}
\label{tab_3fgl_mspyng_fea}
\end{table}

Using these two sets of selected features, we build the prediction models for the AGN/PSR and YNG/MSP classification tasks with various machine learning methods.
A comparison of accuracies between our method and that adopted by  \cite{parkinson2016classification} are presented in Table~\ref{tab_3fgl_agnpsr_accu} and Table~\ref{tab_3fgl_mspyng_accu}.
Both tables show that our method generally achieves a higher nominal accuracy regardless of the method used to build the prediction model, though such difference can be possibly be reconciled with the tolerance of statistical fluctuation reflected by their S. E..
Hence, in terms of the overall accuracy, 
the use of automatic feature selection can lead to a comparable performance in the previous results with a simpler model.

\begin{table}
\centering
\begin{tabular}{c | c | c| c| c}
\hline
\hline
\multicolumn{5}{|c}{AGN/PSR Classification in the 3FGL Catalog}\\
\hline
\multirow{2}{*}{Classifiers} & \multicolumn{2}{|c}{Saz Parkinson et al. (2016)} & \multicolumn{2}{|c}{Our framework} \\
\cline{2-5}
 ~&Accu. & S. E. &Accu. & S. E.\\
\hline
\textbf{Boost LR} & \textbf{97.9$\%$} & {\bf 0.38$\%$} & \textbf{ 98.4$\%$} & {\bf 0.43$\%$}\\
\hline
RF      & 97.7$\%$ & $0.41\%$ & $98.0\%$ & $0.47\%$ \\
\hline
LR      & 97.3$\%$ & $0.52\%$ & $97.7\%$ & $0.56\%$ \\
\hline
SVM     & 97.6$\%$ & $0.50\%$ & 97.7$\%$ & $0.60\%$ \\
\hline
LMT     & 97.3$\%$ & $0.51\%$ & 97.7$\%$ & $0.56\%$ \\
\hline
DT      & 96.8$\%$ & $0.49\%$ & 97.0$\%$ & $0.52\%$ \\
\hline
GAM     & 97.2$\%$ & $0.53\%$ & 97.7$\%$ & $0.58\%$ \\
\hline
\end{tabular}
\caption{The comparison of accuracies (see Equation~\ref{equ_aver_Acc}) and standard error (see Equation~\ref{equ_sd}) between our framework and the approach \citep{parkinson2016classification} for AGN/PSR classification.
The highest classification accuracy is obtained by using our framework with boosted logistic regression (Boost LR)}
\label{tab_3fgl_agnpsr_accu}
\end{table}

\begin{table}
\centering
\begin{tabular}{c | c | c| c| c}
\hline
\hline
\multicolumn{5}{|c}{MSP/YNG Classification in the 3FGL Catalog}\\
\hline
\multirow{2}{*}{Classifiers} & \multicolumn{2}{|c}{Saz Parkinson et al. (2016)} & \multicolumn{2}{|c}{Our framework} \\
\cline{2-5}
 ~&Accu. & S. E. &Accu. & S. E. \\
\hline
\textbf{Boost LR} & \textbf{90.2$\%$} & {\bf 2.16$\%$}  & \textbf{ 93.0$\%$} & {\bf 3.31$\%$}\\
\hline
RF      & 88.2 $\%$ & 2.57$\%$ & 90.7$\%$ & 3.56$\%$ \\
\hline
LR      & 87.0 $\%$ & 2.71$\%$ & 88.7$\%$ & 3.81$\%$ \\
\hline
SVM     & 83.7$\%$ & 4.00$\%$ & 87.4$\%$ & 4.32$\%$ \\
\hline
LMT     & 83.9$\%$ & 2.55$\%$ & 84.3$\%$ & 3.21$\%$ \\
\hline
DT      & 83.5$\%$ & 1.83$\%$ & 84.6$\%$ & 2.98$\%$ \\
\hline
GAM     & 84.8$\%$ & 3.08$\%$ & 86.5$\%$ & 3.04$\%$ \\
\hline
\end{tabular}
\caption{The comparison of accuracies (see Equation~\ref{equ_aver_Acc}) and standard error (see Equation~\ref{equ_sd}) between our framework and the approach \citep{parkinson2016classification} for MSP/YNG classification.
The highest classification accuracy is obtained by using our framework with boosted logistic regression (Boost LR)}
\label{tab_3fgl_mspyng_accu}
\end{table}

One should note that the S. E. of the accuracies for the AGN/PSR classification are generally smaller than 
those for the YNG/MSP classification (see Table~\ref{tab_3fgl_mspyng_accu} and Table~\ref{tab_3fgl_agnpsr_accu}). 
Such difference is stemmed from the different sizes of the training sets in these two tasks. The relatively small sample for training any classifiers in $\gamma-$ray astronomy make us cannot ignore the fluctuations of the accuracy when one quote this performance metric.

\begin{figure*}
\centering
\subfigure[In AGN/PSR classification, the training ROC curve produced by our framework]{
\begin{minipage}[b]{2.5in}
\includegraphics[width=2.5in]{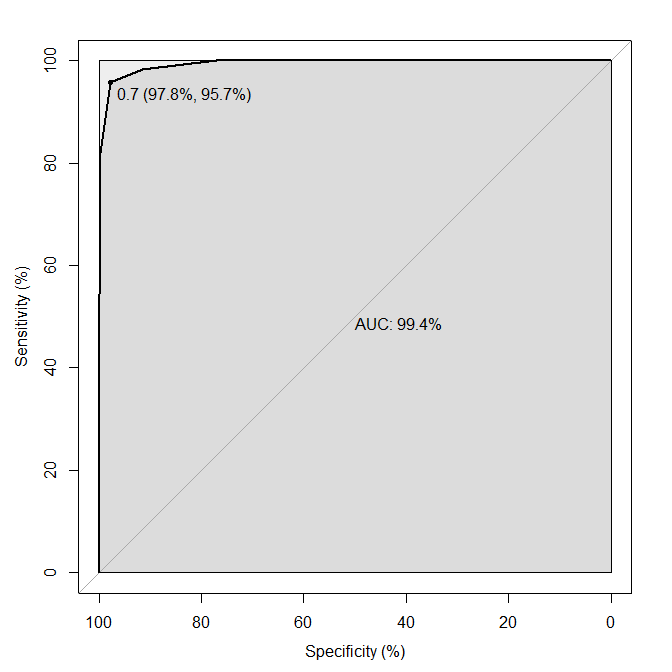} 
\end{minipage}
}
~~~~
\subfigure[In AGN/PSR classification, the training ROC curve produced by the method of Saz Parkinson et al. (2016)]{
\begin{minipage}[b]{2.5in}
\includegraphics[width=2.5in]{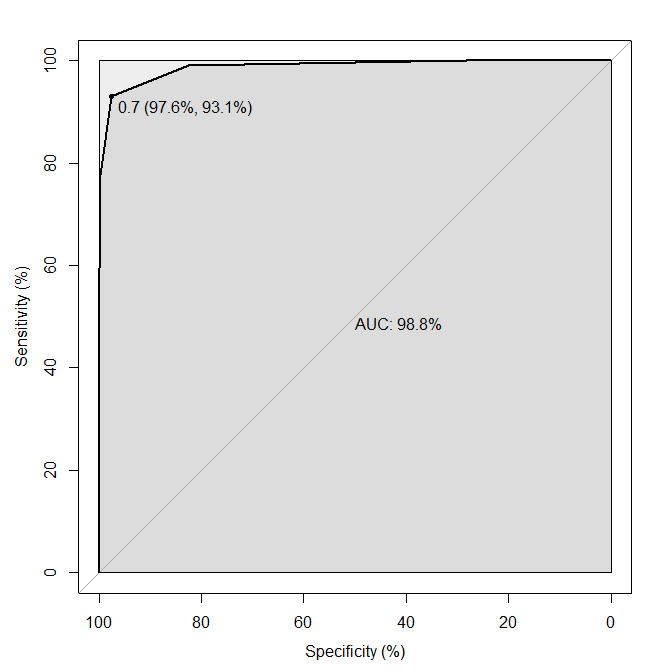} 
\end{minipage}
}
\vfill
\subfigure[In PSR/YNG classification, The training ROC curve produced by our framework]{
\begin{minipage}[b]{2.5in}
\includegraphics[width=2.5in]{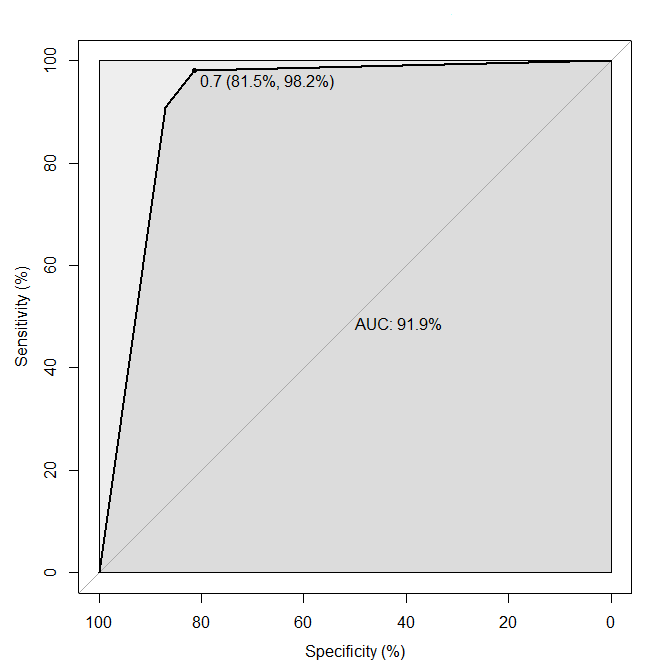} 
\end{minipage}
}
~~~~
\subfigure[In PSR/YNG classification, The training ROC curve produced by the method of Saz Parkinson et al. (2016)]{
\begin{minipage}[b]{2.5in}
\includegraphics[width=2.5in]{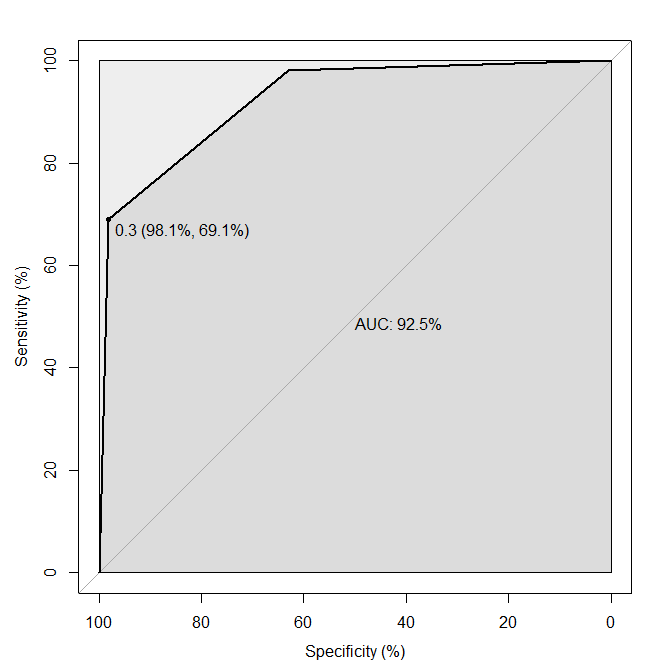} 
\end{minipage}
}
\caption{The training ROC curves of Boosted Logistic Regression (Boost LR) in AGN/PSR and PSR/YNG classification.}
\label{fig_train_roc_blr}
\end{figure*}

\begin{figure*}
\centering
\subfigure[The test ROC produced by Boosted Logistic Regression (Boost LR)]{
\begin{minipage}[b]{3.2in}
\includegraphics[width=2.5in]{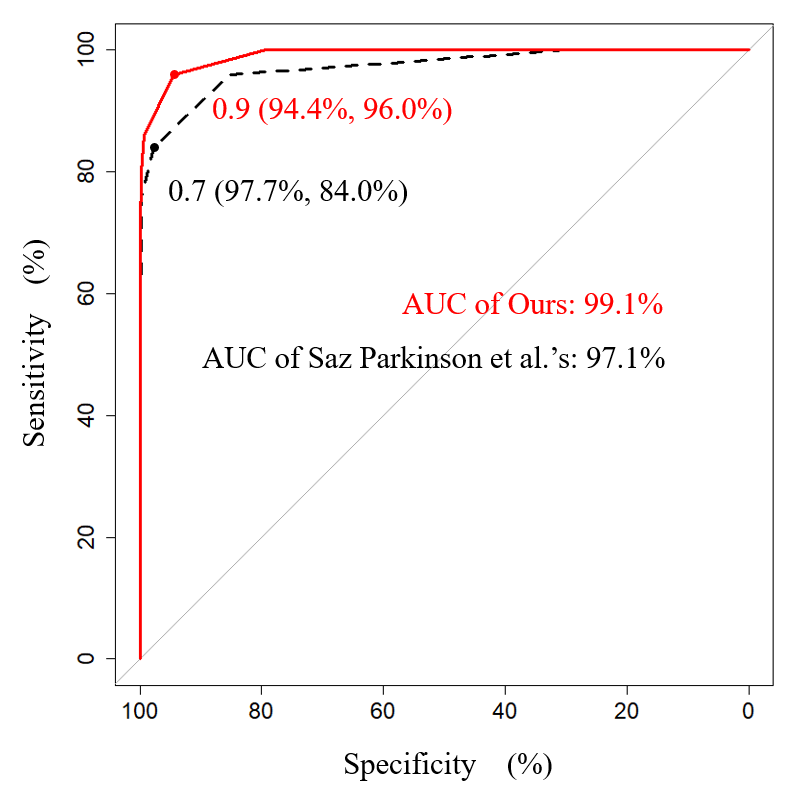} 
\end{minipage}
}
~~~~
\subfigure[The test ROC produced by Boosted Logistic Regression (Boost LR)]{
\begin{minipage}[b]{3.2in}
\includegraphics[width=2.5in]{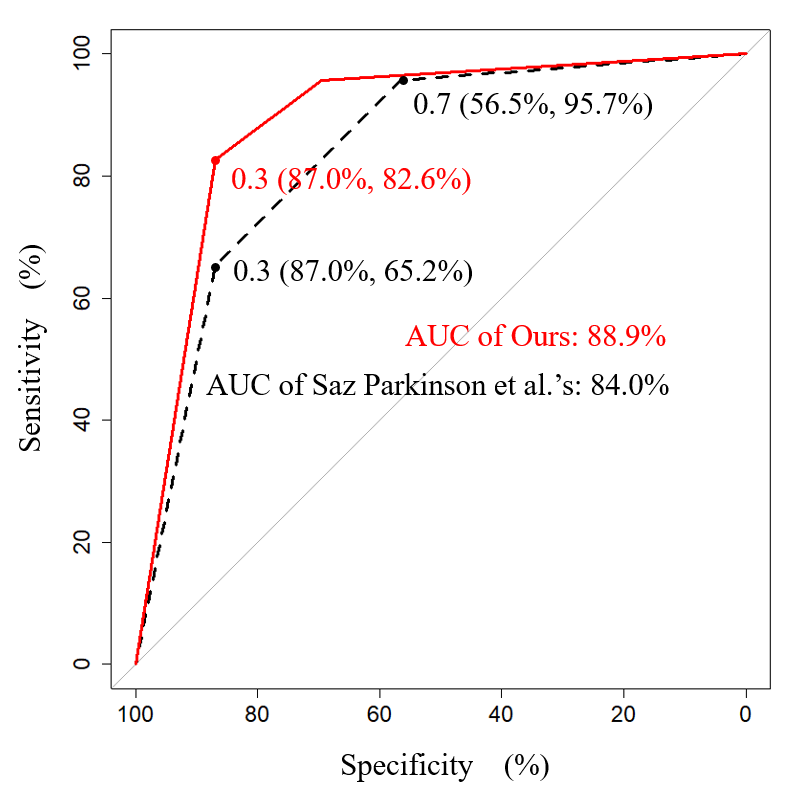}
\end{minipage}
}
\caption{Comparison of the test ROC curves produced by our method and the method of Saz Parkinson et al. (2016) in the AGN/PSR classification and YNG/MSP classification.}
\label{fig_compari_test_roc}
\end{figure*}

Apart from the accuracies (Table~\ref{tab_3fgl_agnpsr_accu} and Table~\ref{tab_3fgl_mspyng_accu}), we also compare the prediction performance of our method and that of \cite{parkinson2016classification} by computing the ROC curves using the test data sets and with the models result in the highest nominal accuracies.
The test ROC curves are shown in Figure~\ref{fig_compari_test_roc} for AGN/PSR and YNG/MSP classifications respectively. On the other hand, the training ROC curves are given in 
Figure~\ref{fig_train_roc_blr}. 

Comparing the test ROC curves produced by our method and that adopted by \cite{parkinson2016classification} (Figure~\ref{fig_compari_test_roc}), our method results in a higher true-positive rate and keep a lower false-positive rate in both classification tasks (i.e. the ROC curve pushed to the top-left corner). And hence,  
the AUC obtained by our method is apparently higher, particularly in the YNG/MSP classification task. 
Together with the higher nominal accuracies obtained by our method, we can observe that by using the set of the features selected by our algorithm (see Algorithm~\ref{alg_rfe_rf})
the classification performance can generally be improved.

\section{Covariate Shift}
\label{sec_covariate_shift}
The usual assumption in supervised learning is that training and test feature vectors are independently
and identically (iid) drawn from the same distribution. When the distributions on the training and the test sets do not match, we have sample selection bias, covariate shift or the Malmquist bias in astronomy. More specifically with sample selection bias, given feature vectors
$X$ and labels $Y$, we have training samples $Z = \{(x_1 , y_1 ), \dots , (x_m , y_m )\} \subset X \times Y$ from a particular probability distribution $Pr(x, y)$, and test $Z' = \{(x'_1 , y'_1 ), \dots , (x'_m , y'_m )\} \subset X \times Y$ drawn from another distribution $Pr'(x, y)$.

For example, in astronomy, the training examples can be biased towards brighter sources as they can be detected easier. 
For variable stars, 
samples of more luminous or well-understood stars are mostly used to train models with supervised learning to classify fainter stars compiled in later and more recent surveys \citep{richards2011active,richards2012overcoming}. This can lead to potential problems as the classification models perform the best with bright objects, not faint objects. Sample selection bias has been ignored in most work in astronomy with machine learning techniques. Especially in astronomy and astrophysics, the problem with bias arises fairly naturally, i.e., with the training data collected in a biased manner (strong sources are more easily found), inevitably the trained model is used to classify objects from a more general target population if sample selection bias is not considered.

In the case of $\gamma$-ray astronomy, samples of previously detected pulsars and AGNs do not necessarily reflect the general population in view of various possible observational biases. However, such possible problem of covariate shift was ignored in all previous studies.

\section{Examining the actual performance in 3FGL/4FGL cross-matched test set}
\label{sec_cross_matched}
We have to stress that the classification model built by any supervised learning techniques must be based on the training sets with labels (i.e. the $\gamma$-sources with identified nature in our case).
Also, all the performance metrics quoted for the aforementioned analysis as well as other previous studies (e.g. \cite{parkinson2016classification}) in classifying $\gamma$-ray sources are based
on the test sets which are constructed by sampling from the pool of identified sources. This raises the question on 
the actual performance of such model in classifying unidentified sources which is  based on the assumption that 
the probability distribution functions of the features are the same as those of the training data. 
However, no one has investigated whether such an assertion is justified.
Motivated by this, in the second part of our experiment, we examine the actual performance of apply such model in classifying unidentified sources. 

\subsection{Constructing the 3FGL/4FGL cross-matched test set}
As there were more dedicated analyses in compiling the 4FGL catalog with eight years LAT data accumulated, some unidentified sources in 3FGL 
catalog now have their natures revealed. 
There are 303 unidentified 3FGL sources which are now confirmed as PSR or AGN in the 4FGL catalog.
These 303 sources allow us to construct a cross-match test set for examining the applicability of the model built by 
the labeled 3FGL sources (i.e. those in Sec.~\ref{sec_3fgl_experiments} and in \cite{parkinson2016classification}) 
in actually classifying the unidentified objects in this catalog.
Given that there are only 31 sources in this cross-match test set can be used for MSP/YNG classification, 
only the AGN/PSR classification will be considered in the experiments of examining possible covariate shift.

For predicting whether a source in the 3FGL/4FGL cross-match test set is a AGN or PSR, classifiers are trained with the labeled sources 
from the 3FGL catalog with our method (i.e Algorithms~\ref{alg_OurMeth} \& \ref{alg_rfe_rf}).
In the 3FGL catalog, there are 1904 identified sources in total can be used for building AGN/PSR classification model.
In order to clearly examining the actual performance in the cross-matched test set, five training sets are constructed, and these training sets contain different number of sources ($x \in \left \{ 60,70,80,90,100 \right \}$\% of the 1904 sources).

\subsection{Results}
Using our method (see Algorithm~\ref{alg_OurMeth}), five models were built with training sets of different sizes as sampled from the same pool of identified AGN and PSR in 3FGL catalog. 
The models built with 70\% of the identified AGN/PSR are consistent with those in Section~\ref{sec_3fgl_experiments}. 
We evaluated their performances by testing with the same cross-matched test set.
All the accuracies (see Equation~\ref{equ_aver_Acc}) and the S.E. (see Equation~\ref{equ_sd}) of our method are shown in Table~\ref{tab_cross_match_accu}.

It is obvious that when the sample size of the training set increases, 
the accuracies of our method are increased and the S.E. are decreased.
When we compare these results with those given in Table~\ref{tab_3fgl_agnpsr_accu} which are based on the test sets 
with identified 3FGL sources, the accuracies of 
classifying the cross-match test set are significantly decreased and the S.E. is increased.
For example, when the Boost LR technique is used as the classifier and 70\% of 1904 sources are used for training, 
the accuracy of our method is $98.4\pm0.4$\% as determined by a test set of identified 3FGL sources (see Table~\ref{tab_3fgl_agnpsr_accu}). However, if we test the same model with the 3FGL/4FGL cross-match test set, we obtain an accuracy of $94.1\pm1.0$\% 
(Table~\ref{tab_cross_match_accu})
Even with the model trained by 100$\%$ of the training data, the drop of accuracy in comparing with Table~\ref{tab_3fgl_agnpsr_accu} 
still can hardly be 
reconciled by statistical fluctuations. Therefore, the classification 
model built with identified $\gamma-$ray sources is likely suffering from the problem of covariate shift when it is applied on 
unidentified objects.  

\begin{table*}
\centering
\begin{tabular}{c | c | c| c| c |c | c | c| c| c | c | }
\hline
\hline
\multicolumn{11}{|c}{AGN/PSR Classification in the 3FGL/4FGL Cross-Matched Test Set}\\
\hline
\multirow{2}{*}{Classifiers} & 
\multicolumn{2}{|c}{60\% Training set} & 
\multicolumn{2}{|c}{70\% Training set} &
\multicolumn{2}{|c}{80\% Training set} &
\multicolumn{2}{|c}{90\% Training set} &
\multicolumn{2}{|c}{100\% Training set} \\
\cline{2-11}
 ~ & Accu. & S. E. & Accu. & S. E. & Accu. & S. E..& Accu. & S. E. & Accu. & S. E.\\
\hline
\textbf{Boost LR} & 
\textbf{93.3$\%$} & {\bf 1.70$\%$} & \textbf{ 94.1$\%$} & {\bf 1.04$\%$} & \textbf{94.9$\%$} & {\bf 1.00$\%$}  & \textbf{95.3$\%$} & {\bf 0.83$\%$} & \textbf{95.6$\%$} & {\bf 0.77$\%$} \\
\hline
RF      &  91.8$\%$ & 2.02$\%$ & 92.6$\%$ & 1.83$\%$ & 92.4$\%$ & 1.61$\%$ & 93.5$\%$ & 1.37$\%$ & 94.1$\%$ & 0.96$\%$ \\
\hline
LR      &  92.6$\%$ & 1.75$\%$ & 93.2$\%$ & 1.46$\%$ & 93.1$\%$ & 1.24$\%$ & 94.1$\%$ & 1.10$\%$ & 93.7$\%$ & 1.21$\%$ \\
\hline
SVM     &  92.0$\%$ & 2.20$\%$ & 92.9$\%$ & 1.50$\%$ & 93.2$\%$ & 1.40$\%$ & 92.9$\%$ & 1.24$\%$ & 92.9$\%$ & 0.95$\%$ \\
\hline
LMT     &  90.0$\%$ & 2.57$\%$ & 92.2$\%$ & 1.21$\%$ & 92.5$\%$ & 1.31$\%$ & 92.0$\%$ &  1.31$\%$ & 93.0$\%$ & 0.74$\%$ \\
\hline
DT      &  91.7$\%$ & 1.88$\%$ & 91.3$\%$ & 1.86$\%$ & 92.0$\%$ & 1.13$\%$ & 92.1$\%$ & 1.52$\%$ & 94.1$\%$ & 1.15$\%$ \\
\hline
GAM     &  92.1$\%$ & 1.90$\%$ & 92.1$\%$ & 1.52$\%$ & 92.4$\%$ & 1.61$\%$ & 93.2$\%$ & 1.40$\%$ & 93.3$\%$ & 1.11$\%$ \\
\hline
\end{tabular}
\caption{For AGN/PSR classification, the summary of overall mean accuracies (Accu.) and mean standard error (S. E.) in the 3FGL/4FGL cross-matched test set. The highest classification accuracy is obtained by using our framework with boosted logistic regression (Boost LR)}
\label{tab_cross_match_accu}
\end{table*}

\subsection{Identifying the features with covariate shift problem}
\begin{figure*}
\centering
\includegraphics[width=7in]{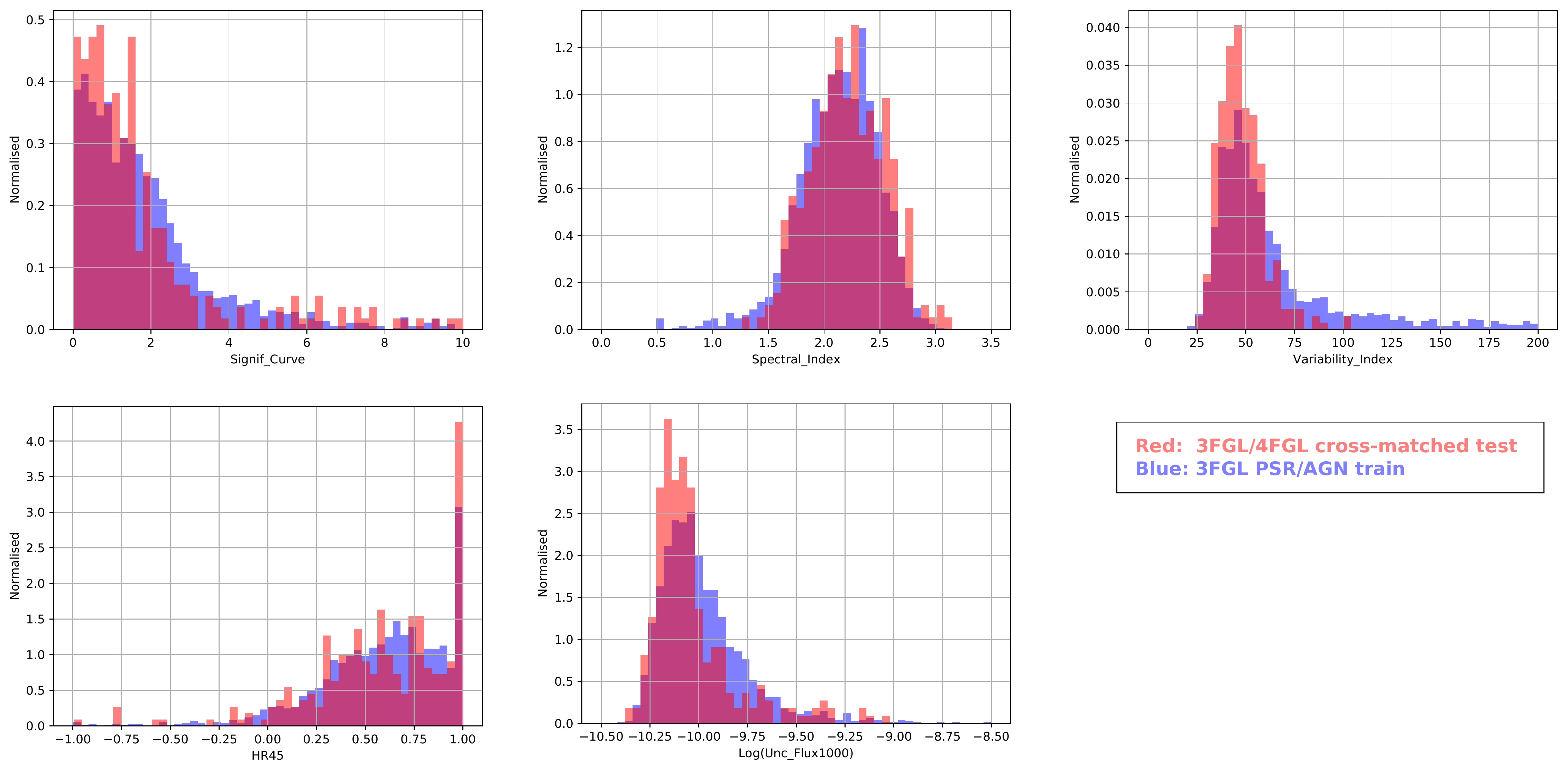}
\caption{Comparisons of the normalized distributions of the features 
selected by RFE between the identified AGN/PSR in 3FGL 
catalog (blue) and the 3FGL/4FGL cross-match test set (red). The purple regions represent the overlaps between 
these training data and the test set.}
\label{3fgl4fgl_agnpsr_comp}
\end{figure*}

Since the problem of the covariate shift has been spotted, we proceeded to find out which feature(s) is/are the 
culprit(s). In Figure~\ref{3fgl4fgl_agnpsr_comp}, we compare the histograms of the five features selected by RFE for 
AGN/PSR classification from the 3FGL training set and the 3FGL/4FGL cross-matched test set. We notice that the 
distributions of some features between these two samples are very different. For quantifying the differences, we
used two-sample Anderson-Darling tests to investigate which feature(s) has/have significant 
covariate shift problem. The results are summarized in Table~\ref{3fgl4fgl_ad}. 

Among all the selected features, the most significant difference between the training data and the cross-match set is 
found in the distribution of \texttt{Variability\_Index} (see Table~\ref{3fgl4fgl_ad}). 
It is obvious that there are significantly more sources with high \texttt{Variability\_Index} 
in the training set (Figure~\ref{3fgl4fgl_agnpsr_comp}). 
Such difference can be a result of selection effect. $\gamma-$ray variability flux is one 
of the defining characteristics of AGNs and it is not difficult to identify their flux variation with the survey mode 
of {\it Fermi} LAT. Because of this, most of the highly variable sources in the 3FGL catalog could be 
readily identified as AGNs. And this can possibly result in a deficit of highly variable sources in the 
pool of unidentified objects. 

While the importance score of \texttt{Variability\_Index} ranks the second among all the features selected by 
our RFE scheme and has been adopted in many other studies in discriminating PSR-like sources from AGN, one has to be 
aware of the covariate shift in the actual classification of the unidentified objects as a result of the aforementioned 
selection effect. The performance 
(e.g. accuracy) can possibly be lower that that determined from a test set sampled from the known sources. 

\begin{figure}
\centering
\includegraphics[width=3in]{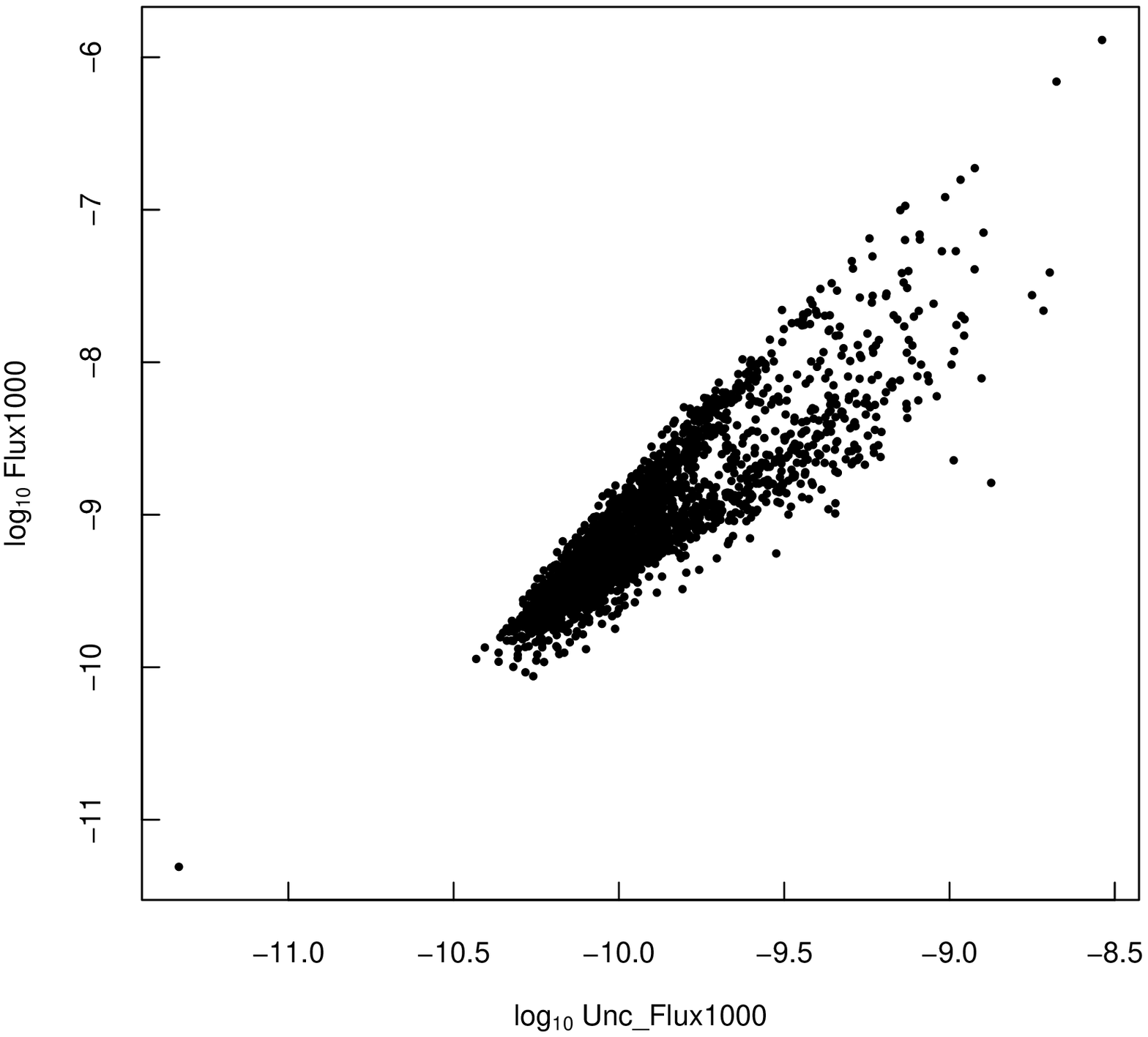}
\caption{The correlation between \texttt{Unc\_Flux1000} and \texttt{Flux1000} in 3FGL catalog.}
\label{3fgl_flux1000}
\end{figure}

\texttt{Unc\_Flux1000} is also found to be significantly different between the training set and the cross-matched 
test set. It appears that the training set contains more sources with higher flux uncertainties 
(see Figure~\ref{3fgl4fgl_agnpsr_comp}). As mentioned in Section~\ref{sec_3fgl_experiments}, 
\texttt{Unc\_Flux1000} can act as a proxy for the source brightness in 1-100 GeV. A source with higher photon 
flux also tends to have a higher flux uncertainty. In Figure~\ref{3fgl_flux1000}, we plot \texttt{Unc\_Flux1000} versus
the photon flux in this band \texttt{Flux1000} for all the sources in 3FGL catalog. It is very clear that 
these two parameters are strongly correlated. This implies that there are more bright sources in the training set.
This can also be a result of selection effect as the bright $\gamma-$ray sources can be analysed in more details (e.g. 
pulsation searches for pinpointing it as a pulsar). 
Also, a brighter $\gamma-$ray source also likely to be bright in the other frequencies so 
as to facilitate multiwavelength identifications.   

The other three selected features are related the spectral shape in $\gamma-$ray. 
\texttt{Signif\_Curve} and \texttt{Spectral\_Index} are based on the results of spectral fitting. 
The differences of their distributions in the training set and the cross-matched test set are much less 
than those of \texttt{Unc\_Flux1000} and \texttt{Variability\_Index}, though they are still not negligible 
(cf. Table~\ref{3fgl4fgl_ad}). \texttt{Signif\_Curve} in 3FGL catalog is obtained by comparing the 
difference of the fitting by a model of power-law with an exponential cutoff and that by a simple 
power-law model (in units of $\sigma$). In order to have a significant discrimination between these two 
models, the sources need to have sufficient photon statistics in the hard band. Therefore, brighter pulsars 
typically have larger \texttt{Signif\_Curve}. This can explain why the training set has more sources with 
large \texttt{Signif\_Curve}. The reason for the excess of objects with small \texttt{Spectral\_Index} (i.e. hard sources) in 
the training data is similar. Sources with small \texttt{Spectral\_Index} are apparently harder in $\gamma-$ray. For a
source with a power-law spectrum to have photons detected in the hard band, it has to be sufficiently bright. 
The relatively small covariate shift of this feature can therefore also be ascribed to the selection effect. 
The excess of hard $\gamma-$ray sources (and hence brighter) can also be seen in the distribution of 
the hardness ratio \texttt{hr45} (cf. Figure~\ref{3fgl4fgl_agnpsr_comp}), 
though there is no significant covariate shift found for this feature (Table~\ref{3fgl4fgl_ad}). 

This analysis of covariate shift in actual $\gamma-$ray source classification echoes to the point that we raised in Section~\ref{sec_OurMethod}.
A simple model enables a easier understanding of the learned model and therefore make the diagnosis of the problem more manageable. 
Also, a model built with fewer features is less likely to encounter the problem of covariate shift.

\begin{table}
\centering
\begin{tabular}{cc}
\hline
\hline
Feature & Null hypothesis probability \\
\hline
Variability\_Index & $5.5\times10^{-27}$ \\
\hline
Unc\_Flux1000 & $5.6\times10^{-9}$  \\
\hline
Signif\_Curve  & $3.4\times10^{-4}$  \\
\hline
Spectral\_Index & $2.0\times10^{-3}$  \\
\hline
hr45  & 0.35 \\
\hline
\end{tabular}
\caption{Summary of the results of A-D tests in comparing the feature distributions between the 3FGL training data and the 3FGL/4FGL cross-match test set.
}
\label{3fgl4fgl_ad}
\end{table}

\section{Identifying Millisecond Pulsar Candidates from 4FGL Unidentified Sources}
\label{sec_4fgl_msp}
With all the aforementioned concerns in applying supervised machine learning techniques in classifying 
$\gamma-$ray sources taking into account, we would like to apply our framework (i.e. Algorithm~\ref{alg_OurMeth}) 
to pick MSP-like sources systematically from the unidentified objects in the 4FGL catalog, 
which has not yet been reported by any other work. 

We started by evaluating the performance of seven machine learning methods in the AGN/PSR and MSP/YNG classifications.
Then, the AGN/PSR and MSP/YNG classifiers with the best performance are coupled into two-layers model to selecting 
MSP-like source from the pool of 4FGL unidentified sources.
Finally, in view of the actual performance of our model can be hampered by the possible covariate shift, 
a threshold cut on the confidence scores has been applied to the MSP-like candidates to reduce the possible misclassifications. 
The final candidate list is given in Table~\ref{msp_candidate}.

\subsection{Developing AGN/PSR and MSP/YNG Classification Models}
In the latest version of 4FGL catalog (released on 15 May 2019), 
there are 3550 sources have been identified/associated as PSRs (248) or AGNs (3302). For the identified PSRs, 
we added a label to denote whether it is a MSP or a YNG according to its rotational period. 
A PSR is labeled as a MSP if its rotational period is $<30$~ms, otherwise it is labeled as a YNG. 
There are 5 PSRs we cannot find any information of their rotational periods and therefore they are excluded in the 
training set for MSP/YNG classification. For the other 243 PSRs, 
we can further divide them into 132 YNGs and 111 MSPs. 
Following the experimental setting in Section~\ref{sec_3fgl_experiments}, 70\% of the 3550 sources are randomly extracted to the training set, and the rest sources are in the test set.

\begin{figure}
\centering
\includegraphics[width=3in,height=2.5in]{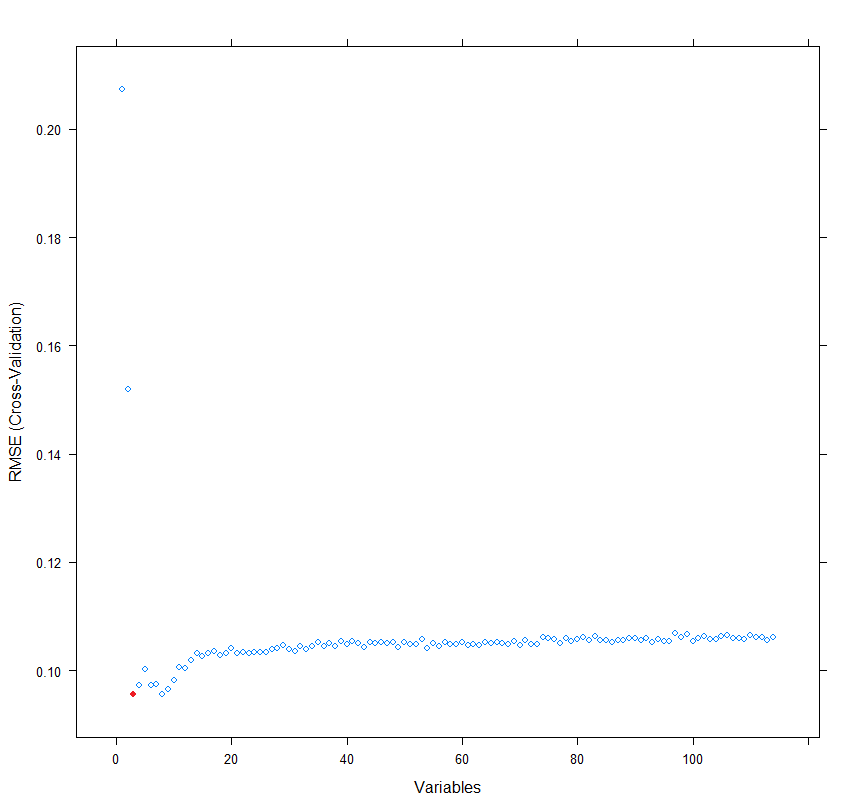}
\caption{The RMSE profile of AGN/PSR classification in the 4FGL catalog. This profile is obtained with the training set. The classifier with the best trade-off between model prediction accuracy and model simplicity is trained with three features (the red point).}
\label{fig_4fgl_profile_agn}
\end{figure}

\begin{figure}
\centering
\includegraphics[width=3in,height=2.5in]{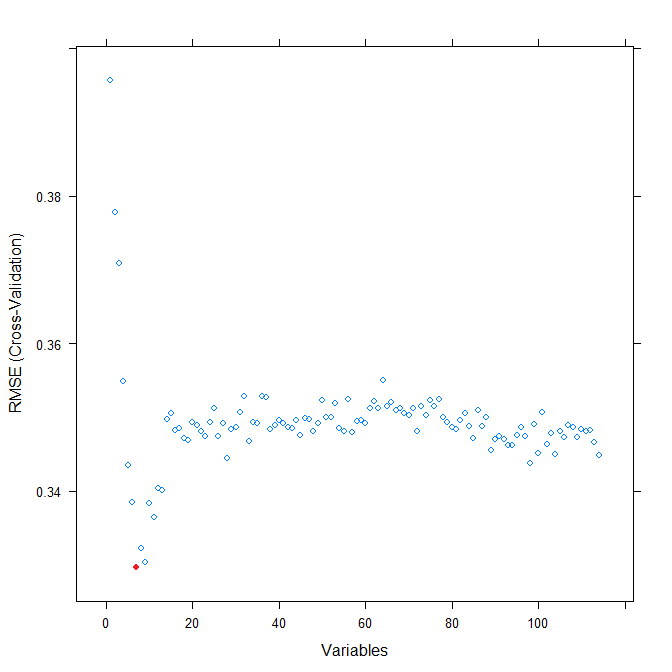}
\caption{The RMSE profile of MSP/YNG classification in the 4FGL catalog. This profile is obtained with the training set. The classifier with the best trade-off between model prediction accuracy and model simplicity is trained with seven features (the red point).}
\label{fig_4fgl_profile_yng}
\end{figure}

In total, there are 114 features in this version of 4FGL catalog available for model building, including 
\texttt{Variability\_Index} which earlier versions of this catalog do not have.
By applying our automatic feature selection method with RFE to the data together with a 1.05 margin of error trade-off 
imposed (see Algorithm~\ref{alg_rfe_rf}), 
three and seven out of 114 features are selected for the AGN/PSR classification and MSP/YNG classification respectively.
The RMSE profiles of both classification tasks are shown in Figure~\ref{fig_4fgl_profile_agn} and Figure~\ref{fig_4fgl_profile_yng}.
We emphasize that these RMSE profiles are obtained with the training set. 
Both sets of selected features with their important scores are shown in Table~\ref{tab_4fgl_agnpsr_fea} and ~\ref{tab_4fgl_mspyng_fea}.

\begin{table}
\centering
\begin{tabular}{ c | c}
\hline
\hline
Features & Importance Scores\\
\hline
PLEC\_SigCurv & 21.90 \\
\hline
LP\_SigCurv & 21.36 \\
\hline
SpectrumType &  19.26\\
\hline
\end{tabular}
\caption{The rank of the features selected by Algorithm~\ref{alg_rfe_rf} for the AGN/PSR classification on the 4FGL training data with labeled sources.
}
\label{tab_4fgl_agnpsr_fea}
\end{table}


\begin{table}
\centering
\begin{tabular}{ c | c}
\hline
\hline
Features & Importance Scores\\
\hline
Unc\_Flux\_Band (10-30 GeV) & 14.87\\
\hline
Unc\_Flux\_Band (3-10 GeV) & 12.43 \\
\hline
GLAT & 9.55 \\
\hline
Unc\_Flux\_Band (1-3 GeV) & 8.97 \\
\hline
LP\_Index & 8.12 \\
\hline
PLEC\_Expfactor & 7.52 \\
\hline
Unc\_Energy\_Flux100 & 7.44 \\
\hline
\end{tabular}
\caption{The rank of the features selected by Algorithm~\ref{alg_rfe_rf} for the MSP/YNG classification on the 4FGL training data with labeled sources.}
\label{tab_4fgl_mspyng_fea}
\end{table}

For the AGN/PSR classification, \texttt{Variability\_Index} is not selected by RFE. This is 
likely due to the fact that the variabilities in this version of 4FGL catalog are computed with the light curves 
over 1-year interval \citep{fermi2019}. Such light curves can be too coarse in discriminating PSRs and AGNs. All three features 
selected for the AGN/PSR classification are all related to the $\gamma-$ray spectral shape. 
\texttt{PLEC\_SigCurv} and \texttt{LP\_SigCurv} replace \texttt{Signif\_Curve} in 3FGL catalog 
in describing how significant the spectra are curved. While \texttt{PLEC\_SigCurv} compares the likelihood of 
a subexponentially cutoff power-law  model with that of a simple power-law model, \texttt{LP\_SigCurv} 
compare that of a log-normal model with a power-law. Another selected feature \texttt{SpectrumType} is
categorical which labels the spectral model of the source as a simple power-law (\texttt{PowerLaw}), 
a subexponentially cutoff power-law (\texttt{PLSuperExpCutoff}) or a log-normal model (\texttt{LogParabola}). 

For YNG/MSP classification, we found that the properties of the features are similar that used in classifying 
YNG/MSP in 3FGL catalog (cf. Table~\ref{tab_3fgl_mspyng_fea}). The uncertainties of the photon flux in three different 
energy ranges and the uncertainty of energy flux in 0.1-100~GeV are chosen. As aforementioned, 
these features can be regarded as proxies for differentiating the brightness between YNGs and MSPs in the corresponding energy bands. 
In 4FGL catalog, the integrated photon fluxes and their uncertainties are given in seven different bands. This can shed light on 
which energy range provide better discrimination between MSPs and YNGs. According to the ranks assigned 
by our algorithm, it appears that hard $\gamma-$ray bands provide better ways to divide these two classes of pulsars. 
Apart from the flux uncertainties, similar to the previous studies, 
the Galactic latitude \texttt{GLAT} is also chosen to discriminate the spatial 
distributions as MSPs tend to populate in the regions further away from the Galactic plane. 

Furthermore, there are more parameters in 4FGL catalog available for separating MSPs from YNGs. 
The other two selected features, namely \texttt{LP\_Index} and \texttt{PLEC\_Expfactor}, describe the 
slope of \texttt{LogParabola} model and the magnitude of the exponential cutoff factor in \texttt{PLSuperExpCutoff} 
model respectively, which cannot be found in the 3FGL catalog.  
These features suggest that the spectral shapes of YNGs and MSPs can be different. Dedicated $\gamma-$ray spectral analysis 
is encouraged to further examine this issue.

\begin{figure*}
\centering
\subfigure[The training ROC curve]{
\begin{minipage}[b]{2.5in}
\includegraphics[width=2.5in]{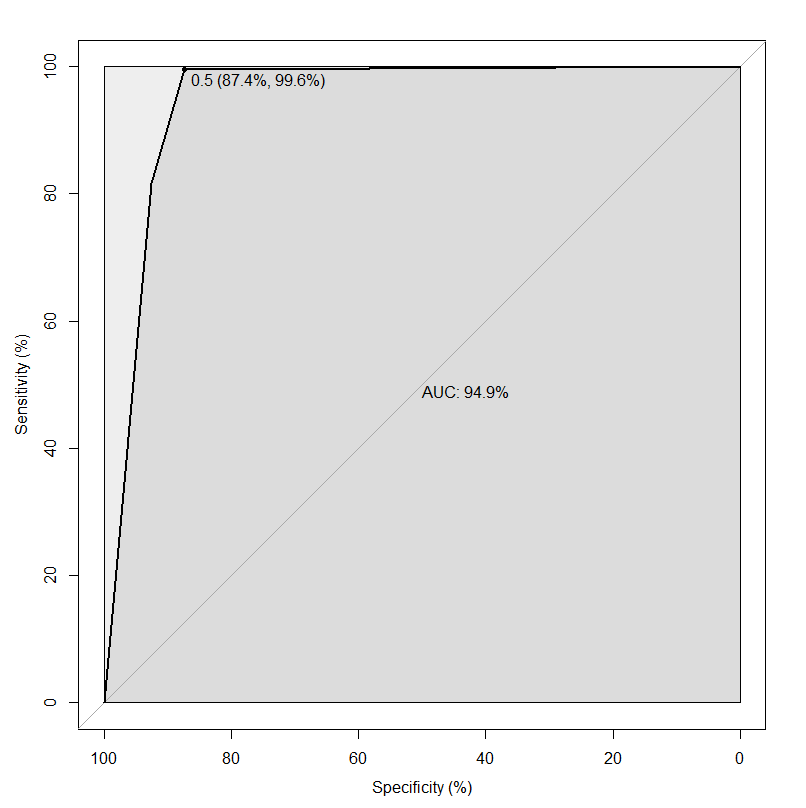} 
\end{minipage}
}
~~~~
\subfigure[The test ROC curve]{
\begin{minipage}[b]{2.5in}
\includegraphics[width=2.5in]{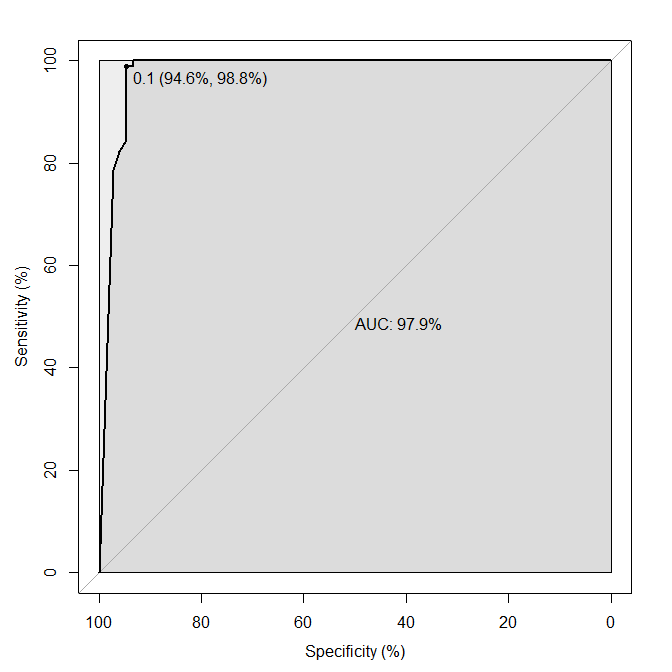} 
\end{minipage}
}
\caption{The training and test ROC curves produced by random forest (RF) classifier using 4FGL catalog in the AGN/PSR classification}
\label{fig_agn_roc_4fhl_rf}
\end{figure*}

\begin{figure*}
\centering
\subfigure[The training ROC curve]{
\begin{minipage}[b]{2.5in}
\includegraphics[width=2.5in]{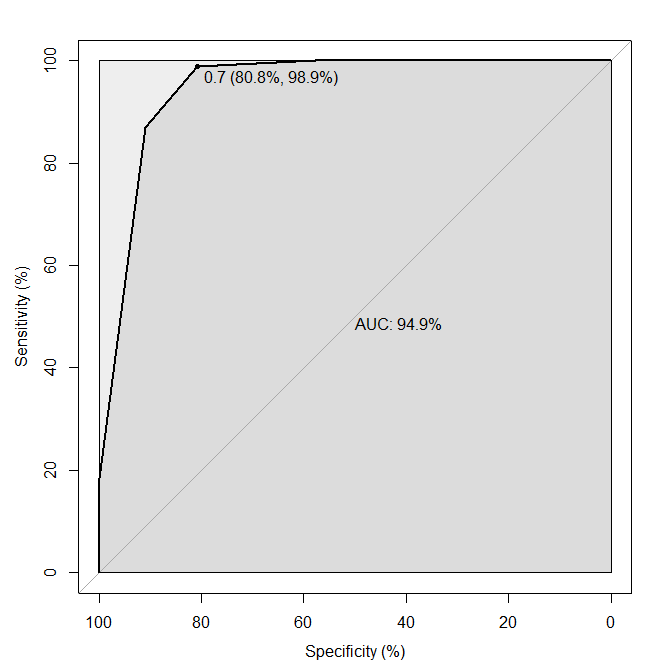} 
\end{minipage}
}
~~~~
\subfigure[The test ROC curve]{
\begin{minipage}[b]{2.5in}
\includegraphics[width=2.5in]{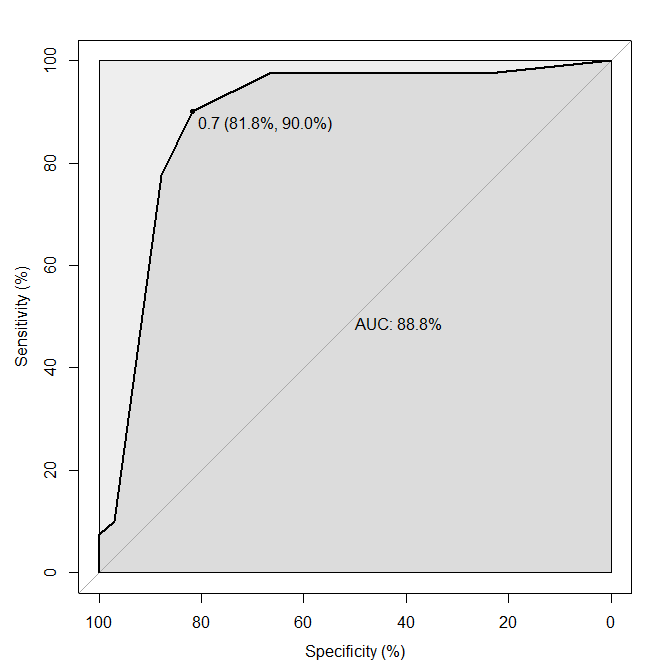} 
\end{minipage}
}
\caption{The training and test ROC curves produced by boosted logistic regression (Boost LR) classifier using 4FGL catalog in the MSP/YNG classification}
\label{fig_msp_roc_4fhl_blr}
\end{figure*}

Using these two sets of selected features, for each classification task, seven classifiers are 
trained based on various machine learning methods.  
The accuracies and their S.E. as computed with Equation~\ref{equ_aver_Acc} and Equation~\ref{equ_sd}
 are summarized in Table~\ref{tab_4fgl_agnpsr_accu} and Table~\ref{tab_4fgl_mspyng_accu} for 
AGN/PSR classification and YNG/MSP classification respectively. 
We show that RF as the classifier attains the highest nominal accuracy of 99.19\% for the 
AGN/PSR classification. Several other classifiers 
(Boost LR, LMT and GAM) also result in the accuracies which are consistent with that of RF within the 
tolerance of the statistical uncertainties, but RF appears to the most stable one as it has 
the lowest S.E.. The training and test ROC curves resulted from this model are shown in 
Figure~\ref{fig_agn_roc_4fhl_rf}.   
For the MSP/YNG classification, the best performance is obtained by the model with Boost LR which has an 
nominal accuracy of 89.61$\%$ with S.E. of 2.34$\%$. 
The corresponding training and test ROC curves are shown in Figure~\ref{fig_msp_roc_4fhl_blr}. 

\subsection{Selection of MSP-like candidates from the unidentified 4FGL sources}
Using two classifiers with the best performance in the AGN/PSR and MSP/YNG classification, we build a 
two-layer cascade method for picking MSP-like sources from the 1154 4FGL unidentified sources.
In the first layer, all 1154 unidentified/unassociated sources are classified into AGN or PSR by using the trained model with RF.
Then, the PSR-like sources selected by the first layer are further classified into MSP or YNG by using the 
Boost LR classifier in the second layer. 

In view of the possible covariate shift when applying this model in classifying real unidentified sources, 
which is difficult to assess in the 4FGL catalog, we apply a selection threshold of pick only those unidentified 
sources with MSP confidence scores $>98\%$ to reduce the chance of misclassifications. 
Finally, 20 MSP-like sources are picked out, and these sources are shown in Table~\ref{msp_candidate}. 
Follow-up multiwavelength observations are encouraged to identify the intrinsic 
nature of these short-listed candidate. 

\section{Summary}
\label{sec_summar}
We have investigated the impacts of several factors on the performance of classifying {\it Fermi} LAT $\gamma-$ray 
sources with supervised learning techniques. 
First, we found that incorporating the classifier with an automatic feature selection algorithm can enable one to 
choose the features for discriminating different classes which can possibly be overlooked by human investigators 
(Section~\ref{sec_feature_selection}). Furthermore, 
this automatic approach can facilitate the updating of a classification model with the inputs of new data. For example, 
in Section~\ref{sec_4fgl_msp}, we have explicitly demonstrated how the AGN/PSR and YNG/MSP classification models can 
be updated from 3FGL catalog to the current version of 4FGL catalog automatically.  

With a small trade-off in the performance (i.e. 5\% in RMSE), we can construct
a simple model with a smaller feature set. 
A simple model can lead us to better understanding of the relationships between the selected 
features and the model predictions instead of using it as black box. Also,  
a simple model is more robust and enables an 
easier diagnosis of any problem (e.g see Section~\ref{sec_cross_matched}).  

In Section~\ref{sec_3fgl_experiments}, we compare various performance metrics of our framework and those based on 
\citet{parkinson2016classification} in AGN/PSR and YNG/MSP classifications with 3FGL catalog. We showed that the 
our prediction models can perform better than \citet{parkinson2016classification} (see Table~\ref{tab_3fgl_agnpsr_accu}, Table\ref{tab_3fgl_mspyng_accu} and Figure~\ref{fig_compari_test_roc}) with 
fewer features. We also pointed out that while the overall accuracy can be a convenient metric for gauging the performance, 
it subjects to non-negligible fluctuations as a result of relatively small samples in $\gamma-$ray astronomy. 
In view of this, we suggest one has to evaluate the accuracy and its standard error with a 10-fold nested cross-validation 
by resampling (Equation~\ref{equ_aver_Acc} and Equation~\ref{equ_sd}). 

The basic assumption of supervised learning is that the training data and the unknown sources that awaited to classified are drawn 
from the same distribution in the feature space. However, the validity of such basic assumption has not be properly examined in all 
previous works in $\gamma-$ray source classifications as the test sets used for evaluating the model were also drawn from 
this identified sources, which come from the same pool for sampling the training data. To examine whether this assumption is 
valid for the AGN/PSR classification model built for 3FGL catalog, we have constructed a test set by cross-matching the unidentified 
sources in 3FGL catalog and the known pulsars or AGNs in 4FGL catalog. Comparing the accuracies based on this cross-match set and 
those reported in Section~\ref{sec_3fgl_experiments} which are based on a test set consists of 3FGL identified sources, we 
have found a significant drop in the performance which can be ascribed to the problem of covariate shift
(Figure~\ref{3fgl4fgl_agnpsr_comp} and Table~\ref{3fgl4fgl_ad}). We found these can be resulted from various selection effect. 
While \texttt{Variability\_Index} have been adopted in many previous work to differentiate PSRs and AGNs, we found that its 
distribution for the identified sources can be seriously biased. 
Since significant flux variation can be readily spotted and this 
is one of the defining properties of AGNs, there are more identified sources have high 
\texttt{Variability\_Index} (Figure~\ref{3fgl4fgl_agnpsr_comp}). The flux uncertainties \texttt{Unc\_Flux1000} also has excess 
at large values in the training data (Figure~\ref{3fgl4fgl_agnpsr_comp}). As \texttt{Unc\_Flux1000} is a proxy of the integrated 
flux (see Figure~\ref{3fgl_flux1000}), brighter sources are more likely to be identified with detailed analysis. Hence, the covariate 
shift of this feature is also a result of selection effect. The other three features selected for building the 3FGL AGN/PSR 
classification model are related to the $\gamma-$ray spectral shape. Although there are also biases found, the difference of their 
distributions between the training set and the cross-match test set are far less significant than those of 
\texttt{Variability\_Index} and \texttt{Unc\_Flux1000} (Table~\ref{3fgl4fgl_ad}). 
This suggest that the spectral shape or the $\gamma-$ray hardness is a less biased factor for discriminating PSR and AGN. 

Lastly, we have applied our framework to update both AGN/PSR and YNG/MSP classification models with 4FGL catalog. 
Using Algorithm~\ref{alg_OurMeth}, such model-updating procedure can be carried out automatically with new features in the 
updated catalog incoporated. In comparison with 3FGL catalog, there are more features included in the 4FGL catalog, including 
fluxes and their uncertainties in seven different energy bands as well as categorical description of the source spectral type. 
These can possibly provide us with a better classification rules. It is interesting to note that it only requires three 
features to achieve an accuracy of $\gtrsim99\%$ for the AGN/PSR classification (Figures~\ref{fig_4fgl_profile_agn}, 
\ref{fig_agn_roc_4fhl_rf} and Table~\ref{sec_summar}). By coupling with YNG/MSP classifier and applying a 
threshold cut on the MSP confidence score, we have selected 20 promising MSP candidates from the pool of 1154 unidentified/unassociated 
4FGL objects. Pulsation searches in different wavelengths are encouraged. Also, for the MSPs belongs to the classes of redbacks and 
black widows, orbital modulations with periods less than one day can also be detected in optical/infrared as a results of pulsar wind 
heating/ellipsoidal modulation as well as in X-ray from the intrabinary shock. Therefore, the candidate list given in 
Table~\ref{msp_candidate} can initiate a series of multiwavelength identification campaign.

\begin{table}
\centering
\begin{tabular}{c | c | c }
\hline
\hline
\multicolumn{3}{|c}{AGN/PSR Classification in the 4FGL Catalog}\\
\hline
 Classifier & Accuracy & Standard Error \\
\hline
Boost LR & 99.17$\%$ & 0.19 \\
\hline
\textbf{RF} &  \textbf{99.19$\%$} & 0.16  \\
\hline
LR      & 94.72$\%$ & 0.30 \\
\hline
SVM     & 95.77$\%$ & 0.31 \\
\hline
LMT     & 99.10$\%$ & 0.20 \\
\hline
DT      & 98.49$\%$ & 0.24 \\
\hline
GAM     & 99.03$\%$ & 0.20 \\
\hline
\end{tabular}
\caption{The overall mean accuracies and standard errors for AGN/PSR Classification in the 4FGL Catalog 
by using our framework.
The highest classification accuracy is obtained by using our framework with random forest (RF).}
\label{tab_4fgl_agnpsr_accu}
\end{table}

\begin{table}
\centering
\begin{tabular}{c | c | c}
\hline
\hline
\multicolumn{3}{|c}{MSP/YNG Classification in the 4FGL Catalog}\\
\hline
 Classifier & Accuracy & Standard Error \\
\hline
\textbf{Boost LR} & \textbf{89.61$\%$} & 2.34 \\
\hline
RF      &  86.30$\%$ & 2.14 \\
\hline
LR      &  83.56$\%$ & 1.82 \\
\hline
SVM     &  84.25$\%$ & 2.76 \\
\hline
LMT     &  84.11$\%$ & 2.68 \\
\hline
DT      &  84.11$\%$ & 5.42 \\
\hline
GAM     &  85.34$\%$ & 2.96 \\
\hline
\end{tabular}
\caption{The overall mean accuracies and standard errors for MSP/YNG Classification in the 4FGL Catalog 
by using our framework.
The highest classification accuracy is obtained by using our framework with boosted logistic regression (Boost LR)}
\label{tab_4fgl_mspyng_accu}
\end{table}

\begin{table}
\centering
\begin{tabular}{l  c c c}
\hline
\hline
4FGL Name & $l$ & $b$ & Significance \\
\hline
          & degree & degree & $\sigma$ \\
\hline
4FGL J0940.3-7610 & 292.2483 & -17.4495 & 18.4 \\
\hline
4FGL J1833.0-3840 & 356.0167 & -13.2500  & 6.5 \\
\hline
4FGL J1922.5-5233 & 344.9273 & -25.8029 & 5.6 \\
\hline
4FGL J2039.4-3616 & 6.3374 & -36.5465 & 12.5 \\
\hline
4FGL J2039.5-5617 & 341.2679 & -37.1465 & 41.8 \\
\hline
4FGL J2112.5-3043 & 14.9039 & -42.4435 & 50.3 \\
\hline
4FGL J0235.3+5650 & 136.8187 & -3.1947 & 10.5 \\
\hline
4FGL J0330.1+5038 & 146.9120 & -4.6964 & 10.1 \\
\hline
4FGL J0736.9-3231 & 246.7860 & -5.5748 & 13.2 \\
\hline
4FGL J0933.8-6232 & 282.2436 & -7.9074 & 38.4 \\
\hline
4FGL J1335.0-5656 & 308.8829 & 5.4196 & 16.7 \\
\hline
4FGL J1400.6-1432 & 326.9817 & 45.0692 & 19.0 \\
\hline
4FGL J1431.0-4432 & 321.0098 & 14.8082 & 7.7 \\
\hline
4FGL J1539.4-3323 & 338.7878 & 17.5321 & 30.5 \\
\hline
4FGL J1544.2-2554 & 344.7576 & 22.6035 & 18.0 \\
\hline
4FGL J1805.1-3618 & 355.6623 & -7.2374 & 12.5 \\
\hline
4FGL J1827.5+1141 & 40.7509 & 10.5546 & 16.0 \\
\hline
4FGL J1842.1+2737 & 57.0481 & 14.0857 & 13.7 \\
\hline
4FGL J2054.2+6904 & 104.3717 & 15.2847 & 12.0 \\
\hline
4FGL J2212.4+0708 & 68.7857 & -38.4799 & 14.1 \\
\hline
\end{tabular}
\label{msp_candidate}
\caption{List of 20 MSP candidates selected from 4FGL unidentified/unassociated object with confidence score >  98\%)}
\end{table}

\section{Future Work}
Our work has shown some factors one need to consider 
when applying machine learning in $\gamma-$ray astronomy. However, there are 
still many other possible approaches one can consider for further improving the performance. 

A number of selected features have missing values in the raw data. 
While \citet{parkinson2016classification} drop all the sources with any missing values 
in their selected features, we have adopted an approach that 
these values are imputed using the mean of the respective feature values in this work. For future analysis, one can consider using median for imputing missing values as it
is robust to outliers.

In this work, RFE is used for feature selection because of its simplicity and acceptable performance. In RFE, a classifier method is wrapped around to select features.
It might incorporate the inductive bias of the wrapped method.
Using filter methods \citep{Sanchez2007,vergara2014,bennasar2015} to solve the above issues can provide a reasonable possibility for extending our work in the future. The filter methods are relatively classifier-agnostic and select features based on how relevant the features are for predicting the target and how redundant they are with one another in the information-theoretic sense. 

Although nested resampling and cross-validations are reasonable evaluation methods, these methods might lead to overestimate the generalization of models.
To make a more complete evaluation of machine learning methods with the limited amount of data available in $\gamma-$ray regime, Leave One Out Cross Validation (LOOCV) \citep{kohavi1995,wong2015,vehtari2017} is worth exploring. 
Also, the techniques of data augmentation can also be considered which provide ways to enrich the dataset by transformations that preserve the class or adding noise.

\section*{Acknowledgements}
\addcontentsline{toc}{section}{Acknowledgements}
CYH is supported by the National Research Foundation of Korea through grant 2016R1A5A1013277 and 2019R1F1A1062071.
Shengda L. and Alex P. L. are funded by the Science and Technology Development Fund, Macau SAR (No. 0019/2018/ASC).
KLL is supported by the Ministry of Science and Technology of Taiwan through grant 108-2112-M-007-025-MY3. 






\bsp	
\label{lastpage}
\end{document}